\PassOptionsToPackage{unicode}{hyperref}
\PassOptionsToPackage{hyphens}{url}
\PassOptionsToPackage{dvipsnames,svgnames,x11names}{xcolor}
\documentclass[
  12pt]{article}
\usepackage{threeparttable}
\usepackage{amsmath,amssymb}
\usepackage{iftex}
\ifPDFTeX
  \usepackage[T1]{fontenc}
  \usepackage[utf8]{inputenc}
  \usepackage{textcomp} 
\else 
  \usepackage{unicode-math}
  \defaultfontfeatures{Scale=MatchLowercase}
  \defaultfontfeatures[\rmfamily]{Ligatures=TeX,Scale=1}
\fi
\usepackage{lmodern}
\ifPDFTeX\else  
\fi
\IfFileExists{upquote.sty}{\usepackage{upquote}}{}
\IfFileExists{microtype.sty}{
  \usepackage[]{microtype}
  \UseMicrotypeSet[protrusion]{basicmath} 
}{}
\makeatletter
\@ifundefined{KOMAClassName}{
  \IfFileExists{parskip.sty}{%
    \usepackage{parskip}
  }{
    \setlength{\parindent}{0pt}
    \setlength{\parskip}{6pt plus 2pt minus 1pt}}
}{
  \KOMAoptions{parskip=half}}
\makeatother
\usepackage{xcolor}
\makeatletter
\ifx\paragraph\undefined\else
  \let\oldparagraph\paragraph
  \renewcommand{\paragraph}{
    \@ifstar
      \xxxParagraphStar
      \xxxParagraphNoStar
  }
  \newcommand{\xxxParagraphStar}[1]{\oldparagraph*{#1}\mbox{}}
  \newcommand{\xxxParagraphNoStar}[1]{\oldparagraph{#1}\mbox{}}
\fi
\ifx\subparagraph\undefined\else
  \let\oldsubparagraph\subparagraph
  \renewcommand{\subparagraph}{
    \@ifstar
      \xxxSubParagraphStar
      \xxxSubParagraphNoStar
  }
  \newcommand{\xxxSubParagraphStar}[1]{\oldsubparagraph*{#1}\mbox{}}
  \newcommand{\xxxSubParagraphNoStar}[1]{\oldsubparagraph{#1}\mbox{}}
\fi
\makeatother

\usepackage{longtable,booktabs,array}
\usepackage{calc} 
\usepackage{etoolbox}
\makeatletter
\patchcmd\longtable{\par}{\if@noskipsec\mbox{}\fi\par}{}{}
\makeatother
\IfFileExists{footnotehyper.sty}{\usepackage{footnotehyper}}{\usepackage{footnote}}
\makesavenoteenv{longtable}
\usepackage{graphicx}
\makeatletter
\def\maxwidth{\ifdim\Gin@nat@width>\linewidth\linewidth\else\Gin@nat@width\fi}
\def\maxheight{\ifdim\Gin@nat@height>\textheight\textheight\else\Gin@nat@height\fi}
\makeatother
\setkeys{Gin}{width=\maxwidth,height=\maxheight,keepaspectratio}
\makeatletter
\def\fps@figure{htbp}
\makeatother

\makeatletter
\@ifpackageloaded{caption}{}{\usepackage{caption}}
\AtBeginDocument{%
\ifdefined\contentsname
  \renewcommand*\contentsname{Table of contents}
\else
  \newcommand\contentsname{Table of contents}
\fi
\ifdefined\listfigurename
  \renewcommand*\listfigurename{List of Figures}
\else
  \newcommand\listfigurename{List of Figures}
\fi
\ifdefined\listtablename
  \renewcommand*\listtablename{List of Tables}
\else
  \newcommand\listtablename{List of Tables}
\fi
\ifdefined\figurename
  \renewcommand*\figurename{Figure}
\else
  \newcommand\figurename{Figure}
\fi
\ifdefined\tablename
  \renewcommand*\tablename{Table}
\else
  \newcommand\tablename{Table}
\fi
}
\@ifpackageloaded{float}{}{\usepackage{float}}
\floatstyle{ruled}
\@ifundefined{c@chapter}{\newfloat{codelisting}{h}{lop}}{\newfloat{codelisting}{h}{lop}[chapter]}
\floatname{codelisting}{Listing}

\makeatother
\makeatletter
\@ifpackageloaded{caption}{}{\usepackage{caption}}
\@ifpackageloaded{subcaption}{}{\usepackage{subcaption}}
\makeatother

\ifLuaTeX
  \usepackage{selnolig}  
\fi
\usepackage[]{natbib}
\usepackage{bookmark}

\IfFileExists{xurl.sty}{\usepackage{xurl}}{} 
\hypersetup{
  hypertexnames=false,
  pdftitle={Rerandomization and high-dimensional covariate adjustment under network interference},
  pdfauthor={Honghao Zhang; Yanghao Wei; Hanzhong Liu},
  pdfkeywords={Approximate neighborhood interference; direct effect; exposure mapping; Lasso; spillover effect},
  colorlinks=true,
  linkcolor={blue},
  filecolor={Maroon},
  citecolor={Blue},
  urlcolor={Blue},
  pdfcreator={LaTeX via pandoc}}

\newcommand{\anon}{1}

\usepackage{amsthm}
\usepackage{cleveref}
\crefname{assumption}{Assumption}{Assumptions}
\Crefname{assumption}{Assumption}{Assumptions}
\usepackage{amsmath,amssymb,amsfonts}
\usepackage{algorithm}
\usepackage{algpseudocode}
\usepackage{bm}
\usepackage{multirow}
\usepackage{comment}

\def\Haj{\textnormal{Haj}}
\def\HT{\textnormal{HT}}
\AtBeginDocument{}

\def\P{\mathbb{P}}

\DeclareMathOperator{\Cov}{Cov}
\DeclareMathOperator{\Var}{Var}
\DeclareMathOperator{\E}{\mathbb{E}}

\newtheorem{assumption}{Assumption}
\newtheorem{theorem}{Theorem}
\newtheorem{proposition}{Proposition}
\newtheorem{lemma}{Lemma}
\newtheorem{remark}{Remark}
\newtheorem{example}{Example}

\newtheorem{corollary}{Corollary}

\begin{document}
\def\spacingset#1{\renewcommand{\baselinestretch}%
{#1}\small\normalsize} \spacingset{1}

\if1\anon
{
  \title{\bf Rerandomization and high-dimensional adjustment with assignment-dependent auxiliary variables under network interference}
  \author{Honghao Zhang, Yanghao Wei and Hanzhong Liu\footnote{We used OpenAI's ChatGPT to assist with proofreading and editing during the preparation of this manuscript. All errors are our own.}\vspace{0.5cm}\\
    Department of Statistics and Data Science, Tsinghua University
    }
  \maketitle
} \fi

\if0\anon
{
  \bigskip
  \bigskip
  \bigskip
  \begin{center}
    {\LARGE\bf Rerandomization and high-dimensional adjustment with assignment-dependent auxiliary variables under network interference}
\end{center}
  \medskip
} \fi

\bigskip
\begin{abstract}

Network experiments are widely used to evaluate interventions in settings such as vaccine trials and social networks, where interference is often present. In these settings, predictive auxiliary variables, such as treatment-weighted neighbor summaries, depend on treatment assignments and generally exhibit nonzero exposure effects. In this paper, we propose rerandomization and covariate adjustment methods based on estimation errors in these auxiliary exposure effects to improve precision. For Mahalanobis distance-based rerandomization, we derive the asymptotic distributions of the Horvitz--Thompson and Hájek estimators under approximate neighborhood interference and quantify the resulting precision gains. At the analysis stage, we propose a covariate adjustment method that acts directly on these estimation errors. This method requires no normalization of the auxiliary variables and is optimal within the corresponding class of linear adjustments. We show that rerandomization and covariate adjustment become asymptotically equivalent as the rerandomization threshold approaches zero, extending this equivalence to settings with interference. For high-dimensional auxiliary variables, we propose Lasso-adjusted estimators and establish their joint asymptotic normality under suitable sparsity conditions. These estimators retain the same asymptotic distribution when combined with fixed-dimensional rerandomization. Simulations and a social-network application illustrate the precision gains from the proposed methods.

\end{abstract}

\noindent%
\newpage
\spacingset{1.8} 

\section{Introduction}

Randomization ensures the validity of causal inference, but chance covariate imbalance in a realized experiment can reduce estimation precision. To mitigate this, rerandomization improves covariate balance at the design stage, whereas regression and covariate adjustment utilize auxiliary information at the analysis stage. The seminal work of \citet{morgan2012rerandomization} for rerandomization and \citet{lin2013agnostic} for regression adjustment established these approaches as standard tools for precision improvement. Subsequent research has clarified how rerandomization constrains, and covariate adjustment removes, predictable randomization variation \citep{li2018asymptotic,li2020rerandomization}, while other work has extended these methods to various experimental designs \citep{li2020rerandomization2K,schochet2022design,lu2023design,wang2023rerandomization,Liu2024,shi2025rerandomization,lu2025rerandomization}. These studies assume the stable unit treatment value assumption (SUTVA).

In network experiments, however, SUTVA may fail. Such experiments are widely used in vaccine trials, online platforms, social networks, and connected communities. In a public health vaccination trial, for example, vaccinating an individual directly protects that individual while also reducing infection risk for their contacts. In these settings, such interference, where one unit's treatment assignment affects the outcomes of other units, is often the primary causal quantity of interest. Researchers thus aim to evaluate a unit's direct treatment effect alongside spillover effects transmitted through network contacts. Literature on exposure mappings and social interactions provides a framework for defining and estimating these effects \citep{manski1993identification,hudgens2008toward,aronow2017estimating,athey2018exact,basse2018analyzing}. More recent work has advanced design-based inference under structured or approximate neighborhood interference (ANI) \citep{leung2022causal,savje2024causal,yu2022estimating,lu2024estimation}, enabling valid inference without relying on parametric outcome models or stochastic network formation models.

Despite these advances, improving estimation efficiency in network experiments introduces distinct difficulties. For fixed baseline covariates, classic Fisher and Lin's regression adjustments may hurt efficiency under network dependence \citep{gao2025causal}. Moreover, in network settings, researchers often observe many local network centralities and treatment-weighted summaries of neighbor behaviors or risk factors alongside baseline covariates. While these auxiliary variables can be predictive for outcomes, they depend on the realized treatment assignment and are not fixed under randomization. As a result, they typically exhibit nonzero exposure effects. Standard rerandomization using the Mahalanobis distance between estimated auxiliary effects and zero, or adjusting these estimates directly, can therefore introduce bias and increase the variance.

Several recent studies examine how auxiliary information can improve inference under network interference. Recognizing the value of assignment-dependent network variables, \citet{lu2024adjusting} propose an optimal linear adjustment for a scalar exposure contrast. However, their method relies on a normalization procedure that varies across different exposure contrasts. \citet{wang2025covariate} establish non-inflation guarantees for covariate adjustment under neighborhood interference within lower-order interaction models \citep{cortez2023exploiting}, while \citet{shi2026gauger} propose a generalized regression adjustment combining flexible prediction with graph-weighted residualization. Beyond analysis-stage adjustments, researchers often wish to enforce design-stage balance on a small subset of particularly important auxiliary variables \citep{junior2025does}. Rerandomization  provides a design-based protection against p-hacking and specification searches that can arise from purely analysis-stage model selection \citep{lu2025rerandomization}. This motivates us to consider whether a simple strategy can combine fixed-dimensional design-stage rerandomization with high-dimensional analysis-stage covariate adjustment, particularly when accommodating multiple exposure contrasts simultaneously.

To address this issue, we propose design-based rerandomization and covariate adjustment methods using auxiliary-effect estimation errors. Although assignment-dependent network variables have nonzero exposure effects, their expected values under randomization are known by design. We define the auxiliary-effect estimation errors as the differences between estimated auxiliary effects and these design-known targets. We show that these estimation errors can be used for both rerandomization and covariate adjustment to guarantee efficiency gains. Our contributions are threefold.

First, we propose a rerandomization scheme tailored for assignment-dependent network variables. Rather than forcing estimated auxiliary effects toward zero, our procedure balances them around their exact targets determined by the randomization mechanism. These design-known targets and the associated covariance matrix are obtained analytically or via Monte Carlo sampling. Under approximate neighborhood interference (ANI), we establish the asymptotic distributions of the Horvitz--Thompson and Hájek estimators under rerandomization. The limiting distributions follow a normal--truncated-normal form, and we quantify the variance reduction from design-stage balance.

Second, we develop a projection-based linear adjustment for multiple exposure-effect contrasts. This method subtracts from the unadjusted estimator its linear projection onto the auxiliary-effect estimation errors. Centering auxiliary estimators at their design-known targets avoids both the parameter-specific normalization required in \citet{lu2024adjusting} and the residualization used in \citet{shi2026gauger}. The estimated coefficient matrix minimizes the asymptotic variance within the class of linear adjustments. Furthermore, we establish a design-analysis duality under interference: rerandomization and regression adjustment rely on the same underlying projection structure, becoming asymptotically equivalent as the rerandomization threshold approaches zero.

Third, we extend the methodology to high-dimensional settings where the number of auxiliary features exceeds the sample size. We construct a row-wise Lasso adjustment using an estimating-equation approach. To handle network dependence, we prove a Bernstein-type concentration inequality for bounded arrays under ANI, establishing the convergence rate of the Lasso coefficients. Building on this bound, we establish the joint asymptotic normality of the Lasso-adjusted estimator and provide valid design-based confidence intervals. Finally, we show that combining fixed-dimensional rerandomization with high-dimensional Lasso adjustment leaves the first-order asymptotic distribution of the final estimator unchanged.

Our analysis is design-based \citep{imbens2015causal,ding2024first}. Potential outcomes, baseline covariates, and the network are treated as fixed, and treatment assignment is the sole source of randomness. The proposed procedures require neither a correctly specified outcome regression model nor a model of network formation.

The remainder of the paper is organized as follows. Section~\ref{sec:setup} introduces the experimental setup, exposure mappings, and causal estimands. Section~\ref{sec:rerandomization} develops rerandomization with network-dependent auxiliary variables and studies its asymptotic properties. Section~\ref{sec:regression-adjustment} presents projection-based adjustment and its relationship with rerandomization. Section~\ref{sec:lasso} develops the high-dimensional Lasso-adjusted estimators. Sections~\ref{sec:Numerical Illustration} and \ref{sec:application} present the simulation results and empirical application, respectively. Section~\ref{sec:conclusions} concludes. The regularity conditions and definitions needed for the theoretical results are included in Sections~\ref{sec:rerandomization} and~\ref{sec:lasso}. Proofs and additional numerical results are omitted from this version.

\section{Framework and notation}\label{sec:setup}

Consider a finite population of $n$ units $\mathcal N_n=\{1,\ldots,n\}$ connected by a fixed, known, undirected network. Let $A=[A_{ij}]_{i,j\in\mathcal N_n}$ denote the adjacency matrix, where $A_{ij}\in\{0,1\}$ indicates an edge between units $i$ and $j$. Let $\mathcal A_n$ denote the collection of admissible networks. For each unit $i$, let $X_i\in\mathbb R^{L_x}$ be a vector of fixed baseline covariates and write $X=[X_i^\top]_{i=1}^n \in \mathbb{R}^{n \times L_x}$. A binary treatment $D_i\in\{0,1\}$ is independently assigned to each unit, although assignment probabilities may differ across units. Let $D=(D_i)_{i\in\mathcal N_n}$. Unit $i$ has potential outcome $Y_i(d)$ under each assignment $d\in\{0,1\}^n$, and we observe $Y_i=Y_i(D)$. Without further restrictions, $Y_i(D)$ may depend on the assignments of all units.

We adopt a design-based perspective: potential outcomes, baseline covariates, and the network are treated as fixed, and all randomness arises from treatment assignment. This perspective stems from the randomization inference principles of \citet{Neyman1923,Fisher1935} and forms the basis of modern causal inference for randomized experiments \citep{lin2013agnostic,li2016evaluation,Liu2020}. Unless otherwise stated, all expectations, variances, and probabilities below are taken with respect to the assignment mechanism.

To summarize relevant treatment configurations, let $T:\mathcal N_n\times\{0,1\}^n\times\mathcal A_n\to\mathcal T$ be a known exposure mapping with finite range $\mathcal T$, and write $T_i=T(i,D,A)$. Exposure mappings provide a low-dimensional representation of network interference and support interpretable comparisons across treatment environments \citep{aronow2017estimating,athey2018exact,basse2018analyzing,savje2024causal}. Let $\pi_i(t)=\mathbb P(T_i=t)$ denote the generalized propensity score and define $\mu_i(t)=\mathbb E\{Y_i(D)\mid T_i=t\}$. The finite-population average is $\mu(t)=n^{-1}\sum_{i=1}^n\mu_i(t)$, and we collect these averages as $\mu=(\mu(t):t\in\mathcal T)$. Following \citet{leung2022causal,savje2024causal}, these quantities remain well defined even if the exposure mapping does not fully capture interference. If the mapping is correctly specified, so that $Y_i(d)=Y_i(d')$ whenever $T(i,d,A)=T(i,d',A)$, then $\mu_i(t)$ reduces to the usual exposure-specific potential outcome.

We consider $F$ exposure effects collected in $\tau=Q\mu=(\tau_1,\ldots,\tau_F)^\top$, where $Q=[q_f^\top]_{f=1}^F$ is a fixed $F\times|\mathcal T|$ contrast matrix with $\sum_{t\in\mathcal T} q_f(t)=0$ for each $f$. We assume that the target contrasts are linearly independent, so $Q$ has full row rank.

\begin{example}\label{ex:2*2}
Consider the exposure mapping $T_i=(D_i,\mathbf 1\{\sum_{j=1}^n A_{ij}D_j>0\})$, which records own treatment and whether unit $i$ has at least one treated neighbor. Ordering the exposure cells as $(0,0),(0,1),(1,0),(1,1)$, the contrast matrix
\[
Q=\frac{1}{2}
\begin{bmatrix}
-1 & -1 & 1 & 1\\
-1 & 1 & -1 & 1\\
1 & -1 & -1 & 1
\end{bmatrix}
\]
defines average direct, spillover, and interaction effects. Such contrasts provide a simple decomposition of own-treatment and neighborhood-treatment effects \citep{hudgens2008toward,leung2022causal}.
\end{example}

We use the Horvitz--Thompson (HT) and H\'ajek estimators as the unadjusted estimators \citep{leung2022causal,gao2025causal}. For each $t\in\mathcal T$, define
\begin{align}
\label{eq1}
\hat\mu_{\HT}(t)
&= \frac{1}{n}\sum_{i=1}^n \frac{\mathbf 1\{T_i=t\}}{\pi_i(t)}Y_i,
\quad \hat\mu_{\Haj}(t) = \frac{\hat\mu_{\HT}(t)}{\hat 1_{\HT}(t)},
\quad
\hat 1_{\HT}(t) = \frac{1}{n}\sum_{i=1}^n \frac{\mathbf 1\{T_i=t\}}{\pi_i(t)}. 
\end{align}
For $*\in\{\HT,\Haj\}$, let $\hat\mu_*=(\hat\mu_*(t):t\in\mathcal T)$ and $\hat\tau_*=Q\hat\mu_*$.  
The generalized propensity scores $\pi_i(t)$ are determined by the assignment mechanism and can be evaluated analytically or approximated by Monte Carlo. Under positivity, the HT estimator is exactly design-unbiased for $\mu(t)$ under the original assignment mechanism, whereas H\'ajek normalization yields a location-invariant estimator of the exposure contrasts. They are consistent and asymptotically normal under the ANI assumption \citep{leung2022causal,gao2025causal}.

\section{Rerandomization}
\label{sec:rerandomization}

\subsection{Balancing with design-known auxiliary effects}

Although both the HT and H\'ajek estimators are asymptotically normal, their efficiency can be improved by appropriately incorporating auxiliary information. In this section, we propose a rerandomization scheme to balance general network-dependent auxiliary variables that are predictive of the outcome. 

Let $G_i=G(i,X,D,A)\in\mathbb R^{L_G}$ denote auxiliary variables that can be computed from baseline covariates, the network, and a candidate assignment before outcomes are observed. The vector $G_i$ may depend on the assignment and need not be a function of the exposure $T_i$. Define $\mu_{G,i}(t)=\E(G_i\mid T_i=t)$, $\mu_G(t)=n^{-1}\sum_{i=1}^n\mu_{G,i}(t)$, and $\mu_G=\operatorname{vec}\{\mu_G(t):t\in\mathcal T\}$. Replacing $Y_i$ by $G_i$ in \eqref{eq1} gives $\hat\mu_{*,G}$ for $*\in\{\HT,\Haj\}$. Let $Q_G=Q\otimes I_{L_G}$, where $I_m$ is the $m\times m$ identity matrix and $\otimes$ denotes the Kronecker product. The population and estimated auxiliary-effect vectors are $\tau_G=Q_G\mu_G$ and $\hat\tau_{*,G}=Q_G\hat\mu_{*,G}\in\mathbb R^{FL_G}$.

To build intuition, first consider the classical case $G_i=X_i$ and $T_i=D_i$, where the auxiliary variables are fixed baseline covariates and the estimand is the direct effect. Because $X_i$ does not depend on treatment assignment, its design-based treatment effect is zero, so $\tau_G=0$. Classical rerandomization therefore seeks assignments for which the realized treatment--control difference in $X_i$ is close to its design target zero. When this covariate difference is correlated with the treatment-effect estimator, reducing its chance variation improves precision \citep{morgan2012rerandomization}. The same principle applies to assignment-dependent auxiliary variables, but the target is no longer generally zero. Because $G_i=G(i,X,D,A)$ changes with the assignment, units under different exposures can have systematically different auxiliary values even under the randomization design. Thus, a nonzero auxiliary exposure effect may be induced by the design itself rather than by chance imbalance. Forcing $\hat\tau_{*,G}$ toward zero would then remove part of this systematic design-induced structure. The appropriate balancing criteria is instead to make the realized auxiliary effect $\hat\tau_{*,G}$ close to its design-known target $\tau_G$, that is, to reduce the estimation error $\hat\tau_{*,G}-\tau_G$.

For HT, $\E(\hat\tau_{\HT,G})=\tau_G$ exactly. For H\'ajek, $\tau_G$ remains the population target although $\E(\hat\tau_{\Haj,G})$ need not equal $\tau_G$ in finite samples. Importantly, $\tau_G$ and the design covariance of $\hat\tau_{*,G}$ require no outcome information and can be calculated analytically or approximated arbitrarily well by design-based Monte Carlo. Following the balance principle of \citet{morgan2012rerandomization}, we therefore accept an assignment when
\begin{equation}\label{eq:balance-rule}
M_*(D,G) =(\hat\tau_{*,G}-\tau_G)^\top \{\Cov(\hat\tau_{*,G})\}^{-1} (\hat\tau_{*,G}-\tau_G) \le a,
\end{equation}
where $a>0$ is prespecified. Thus, rerandomization balances auxiliary-effect estimation errors rather than the auxiliary effects themselves. Smaller $a$ enforces stronger balance but lowers the acceptance probability. In practice, $a$ can be chosen to a target acceptance probability using the design distribution or its asymptotic approximation.

For the Mahalanobis criterion in \eqref{eq:balance-rule}, smaller values of the threshold $a$ correspond to more strict balance and lower acceptance probabilities. A convenient implementation specifies a target acceptance probability $p_a$ and chooses $a$ from the design distribution of $M_*(D,G)$. Under the conditions stated below, $M_*(D,G)$ converges to a $\chi^2_p$ distribution with fixed $p=FL_G$, so $a$ can be chosen as the $p_a$ quantile of this chi-squared distribution. Alternatively, one can simulate many candidate assignments from the known randomization mechanism, compute $M_*(D,G)$ for each assignment, and set $a$ equal to the empirical $p_a$ quantile of the simulated Mahalanobis distances.

The quantities $\tau_G$ and $\Cov(\hat\tau_{*,G})$ are also design-known. When analytic evaluation is inconvenient, they can be approximated using independent design simulations. In particular, averaging simulated HT auxiliary-effect estimators gives an unbiased Monte Carlo estimator of $\tau_G$. For H\'ajek rerandomization, the same population target $\tau_G$ is used even though it need not equal the finite-sample expectation of $\hat\tau_{\Haj,G}$. The asymptotic results in the main text treat these design quantities as known. When Monte Carlo approximations are used, their errors are required to be negligible relative to the stochastic orders entering the corresponding results.

Mahalanobis rerandomization is used for clarity, but the same centering principle applies to more general quadratic criteria. For example, weighted Euclidean or other positive-definite quadratic forms can prioritize selected auxiliary effects \citep{lu2023design,schindl2024unified}. The essential feature is that the criterion acts on $\hat\tau_{*,G}-\tau_G$, rather than on an auxiliary contrast centered at zero.

\subsection{Asymptotic distribution and precision improvement}

In this section, we study the asymptotic properties of the HT and H\'ajek estimators under rerandomization. We assume that the approximate neighborhood interference (ANI) holds. Specifically, let $\ell_A(i,j)$ denote graph distance and $\mathcal N(i,s;A)=\{j:\ell_A(i,j)\le s\}$. Let $D'$ be an independent copy of $D$, and define $D^{(i,s)}$ by retaining $D$ within $\mathcal N(i,s;A)$ and replacing assignments outside this neighborhood by those from $D'$. To control dependence jointly for outcomes and auxiliary variables, define $\theta_{n,s}
=\max_{i\in\mathcal N_n}
\max\left\{
\E|Y_i(D)-Y_i(D^{(i,s)})|,
\,
\E\|G_i(D)-G_i(D^{(i,s)})\|_\infty
\right\}.$
Joint ANI requires $\sup_n\theta_{n,s}\to0$ as $s\to\infty$. This follows the ANI formulation of \citet{leung2022causal}, while additionally controlling the assignment-dependent auxiliary variables needed for joint design-based inference. The remaining conditions impose local exposure dependence, overlap, bounded outcomes and auxiliary variables, sufficiently slow network growth relative to ANI decay, and a nondegenerate limiting covariance. Assumptions~\ref{asu1}--\ref{asu5} below provide the details.   

Let $V_*=[\begin{matrix}V_{*,\tau\tau}&V_{*,\tau G}; \ V_{*,G\tau}&V_{*,GG}\end{matrix}]$ denote the first-order covariance of $\sqrt n(\hat\tau_*-\tau,\hat\tau_{*,G}-\tau_G)^\top$. We now give its construction and the regularity conditions.

\subsubsection{Regularity conditions and covariance construction}\label{app:fixed-conditions}

Let $\ell_A(i,j)$ denote the shortest-path distance, with $\ell_A(i,i)=0$ and $\ell_A(i,j)=\infty$ for disconnected units. Define $\mathcal N(i,s;A)=\{j\in\mathcal N_n:\ell_A(i,j)\le s\}$.

\begin{assumption}[Local exposure mapping]\label{asu1}
There exists a fixed $K\in\mathbb N$ such that, for all $n$, $i$, $d,d'\in\{0,1\}^n$, and $A,A'\in\mathcal A_n$,
$T(i,d,A)=T(i,d',A')$
whenever $\mathcal N(i,K;A)=\mathcal N(i,K;A')$, the induced subnetworks on these neighborhoods agree, and the treatment assignments agree within the neighborhoods.
\end{assumption}

\begin{assumption}[Overlap]\label{asu2}
There exist constants $0<\underline\pi<\bar\pi<1$ such that $\pi_i(t)\in[\underline\pi,\bar\pi]$ for all $n$, $i$, and $t\in\mathcal T$.
\end{assumption}

\begin{assumption}[Bounded outcomes and auxiliary variables]\label{asu3}
There exist finite constants $c_Y$ and $c_G$ such that $|Y_i(d)|\le c_Y$ and $\|G(i,X,d,A)\|_\infty\le c_G$ for all $n$, $i$, and $d\in\{0,1\}^n$.
\end{assumption}

Let $D'$ be an independent copy of $D$ and define $D^{(i,s)}$ by
$D_j^{(i,s)}=D_j$ for $j\in\mathcal N(i,s;A)$ and
$D_j^{(i,s)}=D'_j$ otherwise. Recall that 
\begin{equation}\label{eq:joint-ani}
\theta_{n,s}=\max_{i\in\mathcal N_n}\max\left\{\E|Y_i(D)-Y_i(D^{(i,s)})|,\,\E\|G_i(D)-G_i(D^{(i,s)})\|_\infty\right\}.
\end{equation}

\begin{assumption}[Joint approximate neighborhood interference]\label{asu4}
As $s\to\infty$, $\sup_n\theta_{n,s}\to0$.
\end{assumption}

Assumption~\ref{asu1} requires the exposure mapping to depend only on a fixed-radius neighborhood, as in \citet{leung2022causal,gao2025causal}, while allowing outcomes to depend on more distant assignments under ANI. Assumption~\ref{asu2} is the standard overlap condition used in inverse-probability-weighted exposure-effect estimation \citep{aronow2017estimating,leung2022causal,gao2025causal}. Assumption~\ref{asu3} imposes uniform boundedness of outcomes and auxiliary variables. It is a common regularity condition for network asymptotics and covariate adjustment \citep{leung2022causal,gao2025causal,lu2024adjusting}. Assumption~\ref{asu4} requires the influence of distant assignments on both outcomes and auxiliary variables to decay uniformly. When the exposure mapping correctly captures a fixed-radius interference structure, the outcome component vanishes beyond that radius. If the auxiliary variables are also exact local functions, their component vanishes similarly.

\begin{align*}
Z_{\HT,i}
&=
\begin{pmatrix}
\bigl(
\pi_i(t)^{-1}\mathbf 1\{T_i=t\}Y_i-\mu_i(t)
:t\in\mathcal T
\bigr)\\
\bigl(
\pi_i(t)^{-1}\mathbf 1\{T_i=t\}G_i-\mu_{G,i}(t)
:t\in\mathcal T
\bigr)
\end{pmatrix},\\
Z_{\Haj,i}
&=
\begin{pmatrix}
\bigl(
\pi_i(t)^{-1}\mathbf 1\{T_i=t\}\{Y_i-\mu(t)\}
-\{\mu_i(t)-\mu(t)\}
:t\in\mathcal T
\bigr)\\
\bigl(
\pi_i(t)^{-1}\mathbf 1\{T_i=t\}\{G_i-\mu_G(t)\}
-\{\mu_{G,i}(t)-\mu_G(t)\}
:t\in\mathcal T
\bigr)
\end{pmatrix}.
\end{align*}
Let $\Sigma_*=\Var\{n^{-1/2}\sum_{i=1}^n Z_{*,i}\}$ and $R=\operatorname{diag}(Q,Q_G)$. Define
\begin{equation}\label{eq:V-block}
V_*=R\Sigma_*R^\top=
\begin{bmatrix}
V_{*,\tau\tau} & V_{*,\tau G}\\
V_{*,G\tau} & V_{*,GG}
\end{bmatrix}.
\end{equation}
To state sufficient weak-dependence conditions, define
$M_n(m,k)=n^{-1}\sum_{i=1}^n|\mathcal N(i,m;A)|^k$.
For two unordered pairs of vertices, write
$\ell_A(\{i,k\},\{j,l\})=\min_{u\in\{i,k\},v\in\{j,l\}}\ell_A(u,v)$. Define
\begin{align*}
\mathcal H_n(s,m)
&=
\left\{
(i,j,k,l)\in\mathcal N_n^4:
k\in\mathcal N(i,m;A),\
l\in\mathcal N(j,m;A),\
\ell_A(\{i,k\},\{j,l\})=s
\right\},\\
\mathcal J_n(s,m)
&=
\left\{
(i,j,k,l)\in\mathcal N_n^4:
k\in\mathcal N(i,m;A),\
l\in\mathcal N(j,m;A),\
\ell_A(i,j)=s
\right\}.
\end{align*}
For finite $s\ge0$, let $r^\circ(s)=\max\{\lceil s/2\rceil-1,0\}$ and
\begin{equation}\label{eq:theta-tilde}
\widetilde\theta_{n,s}
=
\theta_{n,r^\circ(s)}
\mathbf 1\{s>2\max(K,1)\}
+
\mathbf 1\{s\le2\max(K,1)\},
\end{equation}
with $\widetilde\theta_{n,\infty}=0$ for disconnected pairs.

\begin{assumption}[Weak dependence]\label{asu5}
There exist $\epsilon_0\in(0,1)$ and integers $m_n\to\infty$ such that the following conditions hold for every fixed nonzero conformable vector $w$. For $*\in\{\HT,\Haj\}$, let
$\sigma_{*,w,n}^2=\Var\{n^{-1/2}\sum_{i=1}^n w^\top Z_{*,i}\}$ and suppose $\sigma_{*,w,n}>0$ for all sufficiently large $n$. Then, $\sigma_{*,w,n}^{-4}n^{-2}
\sum_{s=0}^n
|\mathcal H_n(s,m_n)|
(\widetilde\theta_{n,s})^{1-\epsilon_0}
\to0$, $\sigma_{*,w,n}^{-3}n^{-1/2}M_n(m_n,2)\to0$, and $\sigma_{*,w,n}^{-1}n^{3/2}
(\widetilde\theta_{n,m_n})^{1-\epsilon_0}
\to0$.
In addition, $\Sigma_*\to\Sigma_*^\infty\succ0$.
  \end{assumption}

Assumption~\ref{asu5} balances network growth and interference decay to ensure a central limit theorem for each fixed linear combination of the influence vectors. The three rate conditions respectively control aggregate dependence among local pairs, the growth of local neighborhoods, and the approximation error from truncating distant assignments. The positive-definite limit of $\Sigma_*$ rules out asymptotically degenerate directions. For HT, $V_{\HT}$ is the exact covariance of the corresponding linear representation on the $\sqrt n$ scale. For H\'ajek, the corresponding first-order covariance satisfies $n\Cov\{(\hat\tau_{\Haj}^\top,\hat\tau_{\Haj,G}^\top)^\top\}=V_{\Haj}+o(1)$. Assumptions~\ref{asu1}--\ref{asu5}, together with the full-row-rank condition on $Q$, imply that $V_*$ has a positive-definite finite limit and yield \Cref{CLT} by the Cram\'er--Wold device.

\begin{proposition}\label{CLT}
Under Assumptions~\ref{asu1}--\ref{asu5},
\begin{equation}\label{eq:joint-clt}
V_*^{-1/2}\sqrt n
\begin{pmatrix}
\hat\tau_*-\tau\\
\hat\tau_{*,G}-\tau_G
\end{pmatrix}
\rightsquigarrow
\mathcal N\!\left(0,I_{F(1+L_G)}\right). \nonumber
\end{equation}
\end{proposition}

\citet{gao2025causal} establish the central limit theory (CLT) for the HT and H\'ajek estimators of average potential outcomes and exposure effects under ANI. \Cref{CLT} further accommodates assignment-dependent auxiliary effects needed for rerandomization. Given this proposition, the asymptotic rerandomization theory of \citet{li2018asymptotic} directly yields the conditional limiting distribution of $\hat\tau_*$ under rerandomization in the presence of interference. Write $p=FL_G$, $V^p_{*,\tau\tau}=V_{*,\tau G}V_{*,GG}^{-1}V_{*,G\tau}$, and $V^v_{*,\tau\tau}=V_{*,\tau\tau}-V^p_{*,\tau\tau}$. Let $\varepsilon\sim\mathcal N(0,I_F)$ and $\zeta_{p,a}\overset d=(U\mid U^\top U\le a)$ for $U\sim\mathcal N(0,I_p)$, independently, and define $v_{p,a}=\P(\chi^2_{p+2}\le a)/\P(\chi^2_p\le a)$. We use $\dot\sim$ to denote equality of limiting distributions.

\begin{theorem}\label{asymptotic distribution}
Under Assumptions~\ref{asu1}--\ref{asu5}, for $a\in(0,\infty)$,
\begin{equation}\label{eq:rerand-limit}
\{ \sqrt n(\hat\tau_*-\tau)
\mid M_*(D,G)\le a \}
\ \dot\sim\
(V^v_{*,\tau\tau})^{1/2}\varepsilon
+
V_{*,\tau G}V_{*,GG}^{-1/2}\zeta_{p,a}.
\end{equation}
The corresponding asymptotic covariance is $V_{*,\mathrm{ReM}} = V_{*,\tau\tau} - (1-v_{p,a})V^p_{*,\tau\tau}
\preceq V_{*,\tau\tau}.$
\end{theorem}

Theorem \ref{asymptotic distribution} shows that the asymptotic distribution of $\hat{\tau}_*$ under rerandomization is a convolution of a normal and a truncated-normal component. The normal component $(V_{*,\tau\tau}^v)^{1/2}\varepsilon$ is unaffected by rerandomization, whereas the truncated-normal component $V_{*,\tau G}V_{*,GG}^{-1/2}\zeta_{p,a}$ is more concentrated around zero. Theorem \ref{asymptotic distribution} also demonstrates that rerandomization using Mahalanobis distance never hurts asymptotic efficiency. For effect $f$, the relative asymptotic variance reduction is $(1-v_{p,a})\mathcal R_{*,f}^2$, where $\mathcal R_{*,f}^2=(V^p_{*,\tau\tau})_{ff}/(V_{*,\tau\tau})_{ff}$ is the squared multiple correlation. Thus, more predictive auxiliary effects yield larger precision gains. As the threshold $a\to0$, the asymptotic covariance reduces to ${V}_{*,\tau\tau}^v$ and the asymptotic variance of $\hat\tau_{*,f}$ reduces to $(V_{*,\tau\tau})_{ff}(1-\mathcal R_{*,f}^2)$.

\subsection{Inference}\label{subsec:inference}

\citet{li2020rerandomization2K} develop valid inference for limiting distributions given by convolutions of normal and truncated-normal components. Following their approach, we construct asymptotically valid confidence regions for $\tau$ based on $\hat\tau_*$. By \Cref{asymptotic distribution}, this requires estimating $V^v_{*,\tau\tau}$ and $V^p_{*,\tau\tau}$, and hence $V_*$. Let $K_{ij}(b_n)=\mathbf 1\{\ell_A(i,j)\le b_n\}$ denote the distance-band kernel. The observable counterparts $\hat Z_{*,i}$ replace the unobserved outcome centers with sample analogues while retaining the design-known auxiliary centers.

For the distance-band kernel $K_{ij}(b_n)=\mathbf 1\{\ell_A(i,j)\le b_n\}$, define the estimated influence vectors
\begin{align*}
\hat Z_{\HT,i}
&=
\begin{pmatrix}
\bigl(
\pi_i(t)^{-1}\mathbf 1\{T_i=t\}Y_i-\hat\mu_{\HT}(t)
:t\in\mathcal T
\bigr)\\
\bigl(
\pi_i(t)^{-1}\mathbf 1\{T_i=t\}G_i-\mu_{G,i}(t)
:t\in\mathcal T
\bigr)
\end{pmatrix},\\
\hat Z_{\Haj,i}
&=
\begin{pmatrix}
\bigl(
\pi_i(t)^{-1}\mathbf 1\{T_i=t\}
\{Y_i-\hat\mu_{\Haj}(t)\}
:t\in\mathcal T
\bigr)\\
\bigl(
\pi_i(t)^{-1}\mathbf 1\{T_i=t\}
\{G_i-\mu_G(t)\}
-\{\mu_{G,i}(t)-\mu_G(t)\}
:t\in\mathcal T
\bigr)
\end{pmatrix}.
\end{align*}
 With $R=\operatorname{diag}(Q,Q_G)$, define the network heteroskedasticity-and-autocorrelation-consistent (HAC) covariance estimators \citep{leung2022causal}
\begin{equation}\label{eq:main-hac}
\hat\Sigma_*
=
\frac{1}n
\sum_{i=1}^n\sum_{j=1}^n
K_{ij}(b_n)\hat Z_{*,i}\hat Z_{*,j}^\top,
\quad
\hat V_*=R\hat\Sigma_*R^\top.
\end{equation}

Let $\mathcal N^\partial(i,s;A)=\{j\in\mathcal N_n:\ell_A(i,j)=s\}$ and define
$M_n^\partial(s)=n^{-1}\sum_{i=1}^n|\mathcal N^\partial(i,s;A)|$. Let $\Delta_\mu$ be the $n\times|\mathcal T|$ matrix with row $i$ equal to
$(\mu_i(t)-\mu(t):t\in\mathcal T)$. Let
$K(b_n)=\{K_{ij}(b_n)\}_{i,j=1}^n$, and define
$\mathcal B_*(b_n)=n^{-1}\Delta_\mu^\top K(b_n)\Delta_\mu$.

\begin{assumption}[HAC regularity]\label{asu6}
For some $\epsilon_0\in(0,1)$, (a) $\sum_{s=0}^n M_n^\partial(s)(\widetilde\theta_{n,s})^{1-\epsilon_0}=O(1)$, (b) $M_n(b_n,1)=o(n^{1/2})$, (c) $M_n(b_n,2)=o(n)$, and (d) $\sum_{s=0}^n|\mathcal J_n(s,b_n)|\widetilde\theta_{n,s}=o(n^2)$.
\end{assumption}

Assumption~\ref{asu6} controls network growth, interference decay, and the HAC bandwidth so that sufficiently distant dependence is negligible and the network-HAC covariance estimator is well behaved. Similar conditions are used by \citet{leung2022causal} for network-HAC inference under ANI and by \citet{gao2025causal} in their regression-based analysis of network experiments.

\begin{theorem}\label{covariance estimator}
Under Assumptions~\ref{asu1}--\ref{asu4} and~\ref{asu6},
\begin{align*}
\hat\Sigma_{*,\mu\mu}
&=
\Sigma_{*,\mu\mu}
+
\mathcal B_*(b_n)
+
o_{\P}(1),\\
\hat\Sigma_{*,\mu G}
&=
\Sigma_{*,\mu G}
+
o_{\P}(1),\\
\hat\Sigma_{*,GG}
&=
\Sigma_{*,GG}
+
o_{\P}(1).
\end{align*}
\end{theorem}

As shown in Theorem~\ref{covariance estimator}, the auxiliary-related blocks are consistent: $\hat V_{*,\tau G}=V_{*,\tau G}+o_{\P}(1)$ and $\hat V_{*,GG}=V_{*,GG}+o_{\P}(1)$, whereas $\hat V_{*,\tau\tau}=V_{*,\tau\tau}+Q\mathcal B_*(b_n)Q^\top+o_{\P}(1)$. Here $\mathcal B_*(b_n)=n^{-1}\Delta_\mu^\top K(b_n)\Delta_\mu$, where row $i$ of $\Delta_\mu$ is $(\mu_i(t)-\mu(t):t\in\mathcal T)$.  $\hat V_{*,\tau\tau}$ is not guaranteed to be conservative or positive semidefinite due to finite-population centering bias and kernel choices. The eigenvalue adjustment of \citet{gao2025causal} resolves this issue, ensuring valid conservative inference in practice.

Define $\hat V^p_{*,\tau\tau}=\hat V_{*,\tau G}\hat V_{*,GG}^{-1}\hat V_{*,G\tau}$ and $\hat V^v_{*,\tau\tau}=\hat V_{*,\tau\tau}-\hat V^p_{*,\tau\tau}$. We construct confidence regions using the full plug-in normal--truncated-normal distribution in \Cref{asymptotic distribution}. Specifically, independently generate $\varepsilon\sim\mathcal N(0,I_F)$ and $\zeta_{p,a}\overset d=(U\mid U^\top U\le a)$ with $U\sim\mathcal N(0,I_p)$, and set $\psi_* = n^{-1/2} \{ (\hat V^v_{*,\tau\tau})^{1/2}\varepsilon + \hat V_{*,\tau G}\hat V_{*,GG}^{-1/2}\zeta_{p,a}\}$.
Let $\hat c_{*,1-\alpha}$ be the $(1-\alpha)$ quantile, obtained by Monte Carlo, of
$\psi_*^\top(n^{-1}\hat V^v_{*,\tau\tau})^{-1}\psi_*$. Then, a $(1-\alpha)$ confidence region for $\tau$ is
\begin{equation*}
\left\{ \tau: (\hat\tau_*-\tau)^\top (n^{-1}\hat V^v_{*,\tau\tau})^{-1} (\hat\tau_*-\tau) \le \hat c_{*,1-\alpha} \right\}. 
\end{equation*}
Under Assumptions~\ref{asu1}--\ref{asu6}, if $Q\mathcal B_*(b_n)Q^\top\to\mathcal B_{\tau\tau}^\infty\succeq0$ and the Monte Carlo critical-value approximation is consistent, this region has asymptotic coverage at least $1-\alpha$.

\section{Network-dependent covariate adjustment}
\label{sec:regression-adjustment}

Rerandomization balances network-dependent auxiliary variables at the design stage. Alternatively, auxiliary information can be used for covariate adjustment at the analysis stage. The strategy of \citet{lu2024adjusting} shows that network-dependent auxiliary variables can improve estimation efficiency under interference. However, their approach relies on normalized auxiliary variables for a scalar exposure contrast. In this section, we propose a covariate adjustment based directly on the auxiliary-effect estimation errors introduced in Section~\ref{sec:rerandomization}. This construction avoids parameter-specific normalization, accommodates multiple exposure contrasts jointly, and makes the duality between rerandomization and covariate adjustment transparent.

We motivate the covariate adjustment from a projection perspective. By the joint asymptotic normality in \Cref{CLT}, for any $B\in\mathbb R^{F\times FL_G}$, define $\hat\tau_*(B)=\hat\tau_*-B(\hat\tau_{*,G}-\tau_G)$ and $B_*^{\mathrm{opt}}=V_{*,\tau G}V_{*,GG}^{-1}$. The adjustment removes from $\hat\tau_*$ the component linearly associated with the auxiliary-effect estimation errors, and $B_*^{\mathrm{opt}}$ is the corresponding asymptotic projection coefficient of $\hat\tau_*-\tau$ on $\hat\tau_{*,G}-\tau_G$. Let $V_*^{\mathrm{adj}}(B)$ denote the asymptotic covariance of $\sqrt n{\hat\tau_*(B)-\tau}$. Then, $V_*^{\mathrm{adj}}(B)=V^v_{*,\tau\tau}+(B-B_*^{\mathrm{opt}})V_{*,GG}(B-B_*^{\mathrm{opt}})^\top$. Hence, $B_*^{\mathrm{opt}}$ minimizes asymptotic covariance in the positive-semidefinite order within this linear adjustment class, simultaneously for every linear combination of the target effects.

The oracle coefficient $B_*^{\mathrm{opt}}$ is readily estimated because the required auxiliary covariance blocks are consistently estimable. Specifically, let $\hat B_*^{\mathrm{opt}}=\hat V_{*,\tau G}\hat V_{*,GG}^{-1}$. Under Assumptions~\ref{asu1}--  \ref{asu6}, $\hat B_*^{\mathrm{opt}}$ is consistent for $B_*^{\mathrm{opt}}$. The design-known target $\tau_G$ is equally important: although it need not be zero for assignment-dependent auxiliary variables, it can be calculated analytically or approximated by design-based Monte Carlo without using outcome information. Thus, the feasible covariate-adjusted estimator is $\hat\tau_*(\hat B_*^{\mathrm{opt}})=\hat\tau_*-\hat B_*^{\mathrm{opt}}(\hat\tau_{*,G}-\tau_G).$
This projection representation clarifies the relationship between analysis-stage adjustment and design-stage rerandomization of Section~\ref{sec:rerandomization}. Rerandomization constrains the component of $\hat\tau_*-\tau$ associated with auxiliary-effect estimation errors, whereas covariate adjustment removes its linear projection.

\begin{theorem}\label{duality}
Under Assumptions~\ref{asu1}--\ref{asu6}, for fixed $a>0$,
\begin{align*}
\sqrt n\{\hat\tau_*(\hat B_*^{\mathrm{opt}})-\tau\}
&\ \dot\sim\
(V^v_{*,\tau\tau})^{1/2}\varepsilon,\quad 
\{\sqrt n\{\hat\tau_*(\hat B_*^{\mathrm{opt}})-\tau\}
\mid M_*(D,G)\le a \}
\ \dot\sim\
(V^v_{*,\tau\tau})^{1/2}\varepsilon.
\end{align*}
\end{theorem}

\Cref{duality} shows that the optimally adjusted estimator has the same asymptotic distribution under the original design and under rerandomization based on the same auxiliary information. Thus, once the projection component has been removed by adjustment, conditioning on its Mahalanobis balance does not further change the asymptotic distribution. Conversely, \Cref{asymptotic distribution} shows that rerandomization progressively contracts this same component as $a$ decreases. Its sequential $a\downarrow0$ limit therefore has covariance $V^v_{*,\tau\tau}$, the same covariance attained by optimal adjustment. Together, these results reestablish a design--analysis duality under interference. The analysis of \citet{li2020rerandomization} reveals the corresponding relationship for fixed covariates under no interference. In our setting, the same connection is governed by projections on auxiliary-effect estimation errors. This equivalence is an asymptotic statement and does not imply identical finite-sample performance. Rerandomization may still improve realized balance and yield additional finite-sample gains, as illustrated in Section~\ref{sec:Numerical Illustration}. Moreover, when design-stage balance and analysis-stage adjustment use different variables, rerandomization can retain benefits through components not removed by adjustment. Related work offers useful perspectives on the practical relationship between design- and analysis-stage efficiency improvement \citep{li2020rerandomization,junior2025does}.

Inference for the adjusted estimator is simpler than for the unadjusted estimator under rerandomization because \Cref{duality} yields a Gaussian limit. For any fitted coefficient matrix $\hat B$, estimate its asymptotic covariance by $\hat V_*^{\mathrm{adj}}(\hat B)=(I_F,-\hat B)\hat V_*(I_F,-\hat B)^\top$, which reduces to $\hat V^v_{*,\tau\tau}$ at $\hat B=\hat B_*^{\mathrm{opt}}$. If $Q\mathcal B_*(b_n)Q^\top\to\mathcal B_{\tau\tau}^\infty\succeq0$, an asymptotically conservative $(1-\alpha)$ confidence region for $\tau$ is
\begin{equation}\label{eq:adjusted-confidence-region}
\left\{ \theta: n\{\hat\tau_*(\hat B_*^{\mathrm{opt}})-\theta\}^\top (\hat V^v_{*,\tau\tau})^{-1} \{\hat\tau_*(\hat B_*^{\mathrm{opt}})-\theta\} \le \chi^2_{F,1-\alpha}
\right\}. \nonumber
\end{equation}
Unlike inference for the unadjusted estimator under rerandomization, this Gaussian limit requires no simulation of truncated-normal critical values.

\section{High-dimensional adjustment and inference}
\label{sec:lasso}

Network experiments often collect rich baseline attributes, graph summaries, and treatment-dependent network features, so the dimension $L_G=L_{G,n}$ of the auxiliary vector may grow with $n$. Throughout this section, $F$ and $|\mathcal T|$ remain fixed, while $p_n=FL_G$ may be comparable to or exceed $n$. In the fixed-dimensional setting, the optimal adjustment coefficient for effect $f$, given by row $f$ of $B_*^{\mathrm{opt}}$, solves the population projection equation $V_{*,GG}b_{*,f}=(V_{*,G\tau})_{:,f}$, with solution $b_{*,f}=V_{*,GG}^{-1}(V_{*,G\tau})_{:,f}$. When $p_n$ is large, however, the corresponding sample equation $\hat V_{*,GG}b=(\hat V_{*,G\tau})_{:,f}$ can be unstable or underdetermined because $\hat V_{*,GG}$ may be ill-conditioned or singular. Motivated by the high-dimensional randomization literature \citep{bloniarz2016lasso}, we regularize this sample projection equation using a row-wise Lasso \citep{tibshirani1996regression}. The main additional challenge under network interference is that both the covariance matrix $\hat V_{*,GG}$ and the cross-covariance vector $(\hat V_{*,G\tau})_{:,f}$ are network-dependent and must be estimated accurately. These estimation errors must be accounted for in the asymptotic analysis of the Lasso-adjusted estimator.

\subsection{Row-wise Lasso adjustment}

We retain the covariance notation from Section~\ref{sec:regression-adjustment}, now allowing $V_{*,GG}$ to have a diverging dimension $p_n\times p_n$. For vectors, $\|\cdot\|_q$ denotes the usual $\ell_q$ norm. Let $\operatorname{supp}(b)=\{j:b_j\ne0\}$ denote the support of a vector $b$. For matrices, $\|A\|_{\max}=\max_{j,k}|A_{jk}|$ and $\|B\|_{1,\infty}=\max_f\|b_f\|_1$ when row $f$ of $B$ is $b_f^\top$.

We define the Lasso-adjusted coefficients via a penalized estimating equation approach. For each target effect, we assume that the population projection can be approximated using a sparse coefficient vector. Specifically, for $f=1,\ldots,F$, write $$(V_{*,G\tau})_{:,f}=V_{*,GG}b_{*,f}+u_{*,f},$$ where $b_{*,f}\in\mathbb R^{p_n}$ is sparse and $u_{*,f}$ measures the approximation error. Let $B_*$ have row $b_{*,f}^\top$, and define $s_n=1\vee\max_{1\le f\le F}|\operatorname{supp}(b_{*,f})|$ as the maximal sparsity level, $\kappa_n=1+\max_{1\le f\le F}\|b_{*,f}\|_1$ as the maximal $\ell_1$ size of the projection coefficients, and $\eta_n=\max_{1\le f\le F}\|u_{*,f}\|_\infty$ as the maximal projection approximation error. When $u_{*,f}=0$, $b_{*,f}$ solves the population projection equation exactly; otherwise, the approximation error is controlled jointly with the penalty and other high-dimensional remainder terms.

We estimate each $b_{*,f}$ by replacing the population covariance blocks in its estimating equation with their network-HAC estimates and adding an $\ell_1$ penalty: \begin{equation}\label{eq:hd-row-lasso-main} 
\hat b_{*,f} = \arg\min_{b\in\mathbb R^{p_n}} \left\{ b^\top\hat V_{*,GG}b -2b^\top(\hat V_{*,G\tau})_{:,f} +\lambda_f\|b\|_1 \right\}. \end{equation} Here, $\lambda_f \geq 0$ is a regularization parameter, which can be selected by cross-validation. Without the penalty, the first-order condition is the sample estimating equation $\hat V_{*,GG}b=(\hat V_{*,G\tau})_{:,f}$. The $\ell_1$ penalty regularizes this potentially ill-posed equation and favors sparse solutions. Stacking $\hat b_{*,f}^\top$ gives $\hat B_*^{\mathrm{lasso}}$, and the Lasso-adjusted estimator is $$\hat\tau_*^{\mathrm{lasso}} =\hat\tau_*-\hat B_*^{\mathrm{lasso}}(\hat\tau_{*,G}-\tau_G).$$ The high-dimensional adjustment preserves the projection interpretation of Section~\ref{sec:regression-adjustment}, replacing direct solution of the covariance estimating equation by sparse regularized estimation.

\subsection{Regularity conditions}\label{app:hd-conditions}

In this subsection, we allow the dimension of the auxiliary vector to grow with the sample size. Specifically, $p_n=FL_G$ may diverge as $n\to\infty$.

\subsubsection{Joint ANI and primitive network rates}\label{app:hd-rates}
For a diverging dimension auxiliary vector, we strengthen the ANI requirement so that a common decay sequence controls both outcomes and all auxiliary coordinates. Let $D'$ be an independent copy of $D$, and let $D^{(i,r)}$ agree with $D$ on $\mathcal N(i,r;A)$ and with $D'$ outside this neighborhood.

\begin{assumption}[Joint ANI for outcomes and auxiliary variables]\label{ass:hd-joint-ani}
There is a common nonincreasing sequence $\{\theta_{n,r}:r\ge0\}$ such that $\sup_n\theta_{n,r}\to0$ as $r\to\infty$ and, for every $r\ge0$,
\[
\max_{i\in\mathcal N_n}\E\left|Y_i(D)-Y_i(D^{(i,r)})\right|\le\theta_{n,r},
\quad
\max_{i\in\mathcal N_n}\E\left\|G_i(D)-G_i(D^{(i,r)})\right\|_\infty\le\theta_{n,r}.
\]
\end{assumption}

Assumption~\ref{ass:hd-joint-ani} is the high-dimensional counterpart of Assumption~\ref{asu4}. It implies Assumption~\ref{asu4} and holds automatically under exact finite-range interference for both $Y_i(D)$ and $G_i(D)$. More generally, it allows effects of assignments outside a fixed neighborhood provided their influence decays uniformly over auxiliary coordinates.

Define the maximal neighborhood size $\Delta_n(m)=\max_{i\in\mathcal N_n}|\mathcal N(i,m;A)|$ and the average HAC-neighborhood size $\bar M_n(b_n)=n^{-1}\sum_i|\mathcal N(i,b_n;A)|$. Let $r_n\to\infty$ be deterministic with $r_n\ge K$ eventually. For the pairwise HAC sums, define
\[
\Delta_n^{\mathrm{pair}}(b_n,r_n)=\max_{(i,j):\ell_A(i,j)\le b_n}\left|\left\{(k,l)\in\mathcal N_n^2:\ell_A(k,l)\le b_n,\ \ell_A(\{i,j\},\{k,l\})\le2r_n\right\}\right|.
\]
A simple sufficient bound is $\Delta_n^{\mathrm{pair}}(b_n,r_n)\le\{\Delta_n(b_n+2r_n)\}^2$. Partition the exposure-mean influence vector as $Z_{*,i}=\{(Z_{*,i}^{\mu})^\top,(Z_{*,i}^{\mu_G})^\top\}^\top$, where $Z_{*,i}^{\mu}\in\mathbb R^{|\mathcal T|}$ and $Z_{*,i}^{\mu_G}\in\mathbb R^{|\mathcal T|L_G}$. Define the contrast-level blocks $Z_{*,i}^{\tau}=QZ_{*,i}^{\mu}\in\mathbb R^F$ and $Z_{*,i}^{G}=Q_GZ_{*,i}^{\mu_G}\in\mathbb R^{p_n}$. For $a,b\in\{\tau,G\}$, let
\begin{equation}\label{eq:hd-population-blocks}
V_{*,ab}=\Cov\left(n^{-1/2}\sum_{i=1}^nZ_{*,i}^{a},\ n^{-1/2}\sum_{i=1}^nZ_{*,i}^{b}\right).
\end{equation}
For HT estimator, these are exact covariance blocks on the $\sqrt n$ scale, while for H\'ajek estimator, they are the corresponding asymptotic covariance targets.

For $a,b\in\{\tau,G\}$, define the covariance mass outside the HAC bandwidth by
\[
B_{*,n,ab}(b_n)=\left\|\frac{1}n\sum_{i=1}^n\sum_{j=1}^n\mathbf1\{\ell_A(i,j)>b_n\}\E\{Z_{*,i}^{a}(Z_{*,j}^{b})^\top\}\right\|_{\max},
\]
and let
\[
B_{n,\mathrm{HAC}}(b_n)=\max_{*\in\{\HT,\Haj\}}\left\{B_{*,n,GG}(b_n)\vee B_{*,n,G\tau}(b_n)\vee B_{*,n,\tau G}(b_n)\right\}.
\]
With $\ell_n=\log(ep_n)$, define
\begin{equation}\label{eq:rho-G-def}
\rho_{G,n}=\sqrt{\frac{\Delta_n(2r_n)\ell_n}{n}}+\theta_{n,r_n}.
\end{equation}
For the fixed-dimensional outcome block and the H\'ajek denominator, define $\rho_{\mu,n}=\sqrt{\Delta_n(2r_n)/n}+\theta_{n,r_n}$ and $\rho_{q,n}=\sqrt{\Delta_n(2K)/n}$. The estimated outcome influence vector also contains the finite-population heterogeneity $\mu_i(t)-\mu(t)$. Its contribution to the high-dimensional cross-covariance rate is controlled by
\[
\rho_{H,n}=\sqrt{\frac{\Delta_n(2r_n)M_n(b_n,2)\ell_n}{n}}.
\]
We collect the HAC errors in
\begin{equation}\label{eq:rho-V-def}
\rho_{V,n}=\sqrt{\frac{\bar M_n(b_n)\Delta_n^{\mathrm{pair}}(b_n,r_n)\ell_n}{n}}+\bar M_n(b_n)\theta_{n,r_n}+B_{n,\mathrm{HAC}}(b_n)+\rho_{H,n}+\bar M_n(b_n)\rho_{\mu,n}.
\end{equation}
The first three terms control pairwise stochastic error, ANI localization, and covariance outside the HAC bandwidth. The final two terms arise from estimated outcome centering. No analogous outcome-heterogeneity term enters the $GG$ block because the design-known unit-level auxiliary centers $\mu_{G,i}(t)$ are retained.

\begin{assumption}[High-dimensional HAC regularity]
\label{ass:hd-hac}
The localization radius $r_n$ and HAC bandwidth $b_n$ satisfy $\rho_{G,n}\to0$ and $\rho_{V,n}\to0$.
\end{assumption}

Assumption~\ref{ass:hd-hac} requires the high-dimensional auxiliary imbalance and HAC covariance estimation errors to vanish uniformly. In particular, the condition $\rho_{V,n}\to0$ requires the omitted covariance mass $B_{n,\mathrm{HAC}}(b_n)$ beyond the HAC bandwidth to vanish. When $b_n\to\infty$, a sufficient condition is
\[
\lim_{m\to\infty}\limsup_{n\to\infty}
\sum_{s>m}M_n^\partial(s)
\{\widetilde\theta_{n,s}\}^{1-\epsilon_0}
=0.
\]
Under exact finite-range dependence, a fixed bandwidth is sufficient if it contains the dependence range, in which case the omitted covariance mass is zero.

\subsubsection{Sparse projection and Lasso conditions}\label{app:hd-sparse}
For each target effect, the population projection equation may admit only an approximately sparse solution, as stated in \Cref{ass:hd-sparse-projection} below.

\begin{assumption}[Sparse approximate projection]\label{ass:hd-sparse-projection}
For every $f=1,\ldots,F$, there are $b_{*,f},u_{*,f}\in\mathbb R^{p_n}$ satisfying $(V_{*,G\tau})_{:,f}=V_{*,GG}b_{*,f}+u_{*,f}$. Let $B_*$ have $f$th row $b_{*,f}^\top$, let $S_f=\operatorname{supp}(b_{*,f})$, and define $s_n=1\vee\max_f|S_f|$, $\kappa_n=1+\max_f\|b_{*,f}\|_1$, and $\eta_n=\max_f\|u_{*,f}\|_\infty$. Assume $\eta_n\to0$.
\end{assumption}

Exact sparse projection corresponds to $u_{*,f}=0$. Under approximate projection, $\eta_n$ measures the maximal residual in the sparse population projection equation.

\begin{assumption}[Compatibility]\label{ass:hd-compatibility}
There is a constant $\phi_0>0$ such that, for every $f$ and every $v\in\mathbb R^{p_n}$ satisfying $\|v_{S_f^c}\|_1\le3\|v_{S_f}\|_1$,
\[
v^\top V_{*,GG}v
\ge
\phi_0^2\frac{\|v_{S_f}\|_1^2}{s_n}.
\]
\end{assumption}

Assumption~\ref{ass:hd-compatibility} is the standard compatibility condition on the population auxiliary covariance matrix and ensures sufficient curvature along sparse directions.

\begin{assumption}[Penalty and high-dimensional remainders]
\label{ass:hd-penalty}
Let $\lambda_f>0$ be the theoretical row penalties and $\lambda_n=\max_f\lambda_f$. Write $r_{*,f}=u_{*,f}$ and $\hat r_{*,f}=(\hat V_{*,G\tau})_{:,f}-\hat V_{*,GG}b_{*,f}$, and define
\[
\mathcal E_{\lambda,n}=\bigcap_{f=1}^F\left\{\lambda_f\ge4\|\hat r_{*,f}-r_{*,f}\|_\infty+4\eta_n\right\}.
\]
Assume $\P(\mathcal E_{\lambda,n})\to1$, $\lambda_n\asymp\kappa_n\rho_{V,n}+\eta_n$, $s_n\lambda_n\to0$, and $\sqrt n\,s_n\lambda_n\rho_{G,n}\to0$.
\end{assumption}

The event $\mathcal E_{\lambda,n}$ requires the penalty to dominate the sampling error and approximation error in the projection equation. The condition $s_n\lambda_n\to0$ yields consistency of the Lasso projection coefficients, while $\sqrt n\,s_n\lambda_n\rho_{G,n}\to0$ makes coefficient estimation negligible at the $\sqrt n$ scale. Since $\lambda_n\asymp\kappa_n\rho_{V,n}+\eta_n$, a sufficient condition is
\[
\sqrt n\,s_n(\kappa_n\rho_{V,n}+\eta_n)\rho_{G,n}\to0.
\]
If $\kappa_n=O(1)$, $\eta_n\lesssim\rho_{V,n}$, and $\rho_{G,n}\asymp\rho_{V,n}\asymp\sqrt{\log L_G/n}$, this reduces to $s_n\log L_G/\sqrt n\to0$. Under network dependence, $\rho_{G,n}$ and $\rho_{V,n}$ additionally record neighborhood growth, ANI decay, local-pair dependence, and the HAC bandwidth.

\begin{assumption}[Additional conditions for covariance estimation]
\label{ass:hd-variance}
Assume
\[
\|V_{*,GG}\|_{\max}+\|V_{*,G\tau}\|_{\max}=O(1), \quad \kappa_n^2\rho_{V,n}\to0, \quad \kappa_ns_n\lambda_n\to0.
\]
\end{assumption}

Assumption~\ref{ass:hd-variance} controls the additional remainder terms introduced by estimating the high-dimensional projection coefficients in the plug-in covariance estimator. The bounded covariance condition prevents the population blocks from diverging, while the two rate conditions make the resulting linear and quadratic estimation errors asymptotically negligible.

\begin{remark}
If \(u_{*,f}=0\), Assumption~\ref{ass:hd-sparse-projection} reduces to exact sparse projection. If one wants to use the \(\ell_2\)-approximation formulation
\[
    \max_f\|u_{*,f}\|_2\le \delta_n n^{-1/4},
    \quad \delta_n\to0,
\]
then \(\eta_n\le\delta_n n^{-1/4}\). The CLT remainder below further requires $\sqrt n\,s_n\eta_n\rho_{G,n}\to0.$
In the benchmark case \(\rho_{G,n}\asymp\sqrt{\log L_G/n}\), this becomes $s_n\delta_n n^{-1/4}\sqrt{\log L_G}\to0.$
Thus, the approximation residual must be small enough relative to the high-dimensional imbalance.
\end{remark}

\subsubsection{Projected CLT conditions}\label{app:projected-clt}
Define the oracle-adjusted influence vector $e_{*,i,n}=Z_{*,i}^{\tau}-B_*Z_{*,i}^{G}\in\mathbb R^F$, whose covariance is $V_*^{\mathrm{adj}}(B_*)$. Let $q\ge0$ be fixed. When $q>0$, let $S_n\in\mathbb R^{q\times p_n}$ be a coordinate-selection matrix whose rows are distinct standard basis vectors of $\mathbb R^{p_n}$ and select the auxiliary contrasts used for rerandomization. When $q=0$, take $S_n$ to be the empty matrix. Define
\begin{equation}\label{eq:hd-unified-projected-influence}
\xi_{*,i,n}^{H}=\begin{pmatrix}Z_{*,i}^{\tau}-B_*Z_{*,i}^{G}\\S_nZ_{*,i}^{G}\end{pmatrix}\in\mathbb R^{F+q},
\end{equation}
with the second block omitted when $q=0$.

\begin{assumption}[Projected CLT for Lasso adjustment and rerandomization]\label{ass:hd-oracle-clt}
The boundedness, ANI localization, weak-dependence, and nondegenerate covariance conditions in Assumptions~\ref{asu3}--\ref{asu5} hold for the fixed-dimensional array $\{\xi_{*,i,n}^{H}:i\in\mathcal N_n\}$.
  \end{assumption}

Assumption~\ref{ass:hd-oracle-clt} covers both ordinary Lasso adjustment and its combination with low-dimensional rerandomization. When $q=0$, it gives the projected CLT for the $F$-dimensional oracle-adjusted error. For fixed $q>0$, recall that $\tau_H=S_n\tau_G$ and $\hat\tau_{*,H}-\tau_H=S_n(\hat\tau_{*,G}-\tau_G)$. The corresponding covariance blocks are $V_{*,\tau H}=V_{*,\tau G}S_n^\top$, $V_{*,GH}=V_{*,GG}S_n^\top$, and $V_{*,HH}=S_nV_{*,GG}S_n^\top$. Let $U_*=(u_{*,1},\ldots,u_{*,F})\in\mathbb R^{p_n\times F}$. Assumption~\ref{ass:hd-sparse-projection} implies
\begin{equation}\label{eq:hd-rem-nested-orthogonality}
V_{*,\tau H}-B_*V_{*,GH}=(V_{*,\tau G}-B_*V_{*,GG})S_n^\top=U_*^\top S_n^\top, \quad \|U_*^\top S_n^\top\|_{\max}\le\eta_n.
\end{equation}
Thus, the oracle-adjusted error is asymptotically orthogonal to the selected balance vector. Assumption~\ref{ass:hd-oracle-clt} then provides the joint fixed-dimensional CLT for their linearized influence representation. For the H\'ajek estimator, the corresponding ratio-linearization remainders are negligible under Assumptions~\ref{ass:hd-hac}--\ref{ass:hd-penalty}.

\subsection{Concentration and high-dimensional error rates}

To establish the convergence rate of the Lasso coefficients and the asymptotic normality of the Lasso-adjusted estimators, we first control two high-dimensional quantities: the auxiliary-effect estimation error $\hat\tau_{*,G}-\tau_G$ and the covariance estimation errors entering the sample projection equation \eqref{eq:hd-row-lasso-main}. These two problems involve different forms of network dependence. The former is a sum of unit-level contributions, whereas the latter is based on local pairs through the network-HAC estimator. We therefore establish separate concentration results for the two quantities and use them jointly in the subsequent Lasso analysis.

Two features of the interference structure are central to the concentration analysis. We require the exposure mapping to depend only on assignments and network information within a fixed radius $K$, so for any $r\ge K$, the indicator $\mathbf 1\{T_i=t\}$ is measurable with respect to assignments in $\mathcal N(i,r;A)$ (see Assumption~\ref{asu1}). We also impose a common decay sequence $\{\theta_{n,r}\}$ controlling the effect of assignments outside $\mathcal N(i,r;A)$ on the outcome and all coordinates of the auxiliary vector (see Assumption~\ref{ass:hd-joint-ani}). These conditions allow the contributions to the HT and H\'ajek estimators to be approximated by variables depending only on local assignments. We first establish a Bernstein concentration inequality for bounded arrays admitting such local approximations.

\begin{lemma}
\label{thm:ani-concentration}
Suppose the treatment assignments are independent. For each $n$, let $\{U_{n,i}\in\mathbb R^{d_n}:i\in\mathcal N_n\}$ be centered with $\max_i\|U_{n,i}\|_\infty\le M$ almost surely, where $M<\infty$ does not depend on $n$. Suppose that, for every integer $r\ge0$, there exist centered approximations $U_{n,i}^{(r)}$, measurable with respect to $\sigma\{D_j:j\in\mathcal N(i,r;A)\}$, such that $\max_i\E\|U_{n,i}-U_{n,i}^{(r)}\|_\infty\le\theta_{n,r}$ and $\max_i\|U_{n,i}^{(r)}\|_\infty\le M$ almost surely. Let $\Delta_n(r)=\max_i|\mathcal N(i,r;A)|$. Then there exists a universal constant $C<\infty$ such that, for every $r\ge0$ and $\delta,\eta\in(0,1)$,
\begin{equation}\label{eq:ani-concentration}
\P\bigg[ \Big\|\frac{1}n\sum_{i=1}^nU_{n,i}\Big\|_\infty > C\bigg\{ M\sqrt{\frac{\Delta_n(2r)\log(2d_n/\delta)}{n}} + M\frac{\Delta_n(2r)\log(2d_n/\delta)}{n} + \frac{\theta_{n,r}}{\eta} \bigg\} \bigg] \le\delta+\eta.
\end{equation} 
\end{lemma}

\Cref{thm:ani-concentration} separates two sources of error. After localization, $U_{n,i}^{(r)}$ and $U_{n,j}^{(r)}$ are independent whenever their $r$-neighborhoods are disjoint, so the localized variables admit a dependency graph controlled by $\Delta_n(2r)$. The first two terms in \eqref{eq:ani-concentration} therefore capture stochastic fluctuation under local dependence, while the final term controls the approximation error from assignments outside the $r$-neighborhood. The proof combines this localization argument with the dependency-graph Bernstein inequality of \href{https://doi.org/10.1002/rsa.20008}{Janson (2004)}. 

Next, we apply \Cref{thm:ani-concentration} to the auxiliary-effect estimators. For a deterministic sequence $r_n\to\infty$ with $r_n\ge K$, let $\ell_n=\log(ep_n)$ and define $\rho_{G,n}=\{\Delta_n(2r_n)\ell_n/n\}^{1/2}+\theta_{n,r_n}$.

\begin{corollary}
\label{cor:hd-tauG-concentration}
Under Assumptions~\ref{asu1}--\ref{asu3} and \ref{ass:hd-joint-ani}--\ref{ass:hd-hac}, for $*\in\{\HT,\Haj\}$, $\|\hat\tau_{*,G}-\tau_G\|_\infty=O_{\P}(\rho_{G,n}).$
\end{corollary}

\Cref{cor:hd-tauG-concentration} shows that all auxiliary-effect estimation errors are uniformly controlled at rate $\rho_{G,n}$ even when their dimension $p_n$ diverges. This max-norm bound is essential for high-dimensional adjustment because the error from estimating the projection coefficients is multiplied by $\hat\tau_{*,G}-\tau_G$ in the Lasso-adjusted estimator. Together with the coefficient convergence rate established below, it determines whether estimating the high-dimensional projection is negligible on the $\sqrt n$ scale.

The covariance blocks entering the Lasso estimating equation require a different argument because the network-HAC estimator is a sum over pairs of nearby units. Its error therefore depends not only on unit-level localization but also on local-pair dependence, covariance beyond the HAC bandwidth, and the replacement of unobserved outcome centers by sample analogues. We collect these contributions in the rate $\rho_{V,n}$ defined in Equation \eqref{eq:rho-V-def}.
\begin{proposition}
\label{prop:hd-cov-concentration}
Under Assumptions~\ref{asu1}--\ref{asu3}, \ref{ass:hd-joint-ani}, and \ref{ass:hd-hac},
\begin{equation}\label{eq:hd-cov-rate-main}
\max\{\|\hat V_{*,GG}-V_{*,GG}\|_{\max}, 
\|\hat V_{*,G\tau}-V_{*,G\tau}\|_{\max} \}
=
O_{\P}(\rho_{V,n}),
\quad *\in\{\HT,\Haj\}. \nonumber
\end{equation} 
\end{proposition}
\Cref{prop:hd-cov-concentration} shows that the high-dimensional covariance blocks entering the Lasso estimating equation can be estimated uniformly in max norm at rate $\rho_{V,n}$. In particular, both the Gram matrix $\hat V_{*,GG}$ and the cross-covariance vector $\hat V_{*,G\tau}$ are sufficiently close to their population counterparts. They support consistent estimation of the sparse projection coefficients. This result, combined with \Cref{cor:hd-tauG-concentration}, yields the convergence rate of the Lasso coefficients and establishes that their estimation error is asymptotically negligible for the Lasso-adjusted exposure-effect estimator. The covariance rate differs from the auxiliary-effect rate because the network-HAC estimator aggregates contributions over local pairs rather than individual units. Its analysis must control pairwise dependence, covariance beyond the HAC bandwidth, and feasible outcome centering in addition to ANI localization. Consequently, $\rho_{V,n}$ and $\rho_{G,n}$ need not have the same order.

\subsection{Asymptotic normality and inference}

\Cref{cor:hd-tauG-concentration} and \Cref{prop:hd-cov-concentration} provide the two concentration results for the Lasso analysis. For a sparse approximate solution $b_{*,f}$ to the population projection equation, \Cref{prop:hd-cov-concentration} implies a max-norm residual of order $O_{\P}(\kappa_n\rho_{V,n}+\eta_n)$ in the sample analogue, motivating $\lambda_n=\max_f\lambda_f\asymp\kappa_n\rho_{V,n}+\eta_n$. Under Assumption~\ref{ass:hd-compatibility}, $\|\hat B_*^{\mathrm{lasso}}-B_*\|_{1,\infty}=O_{\P}(s_n\lambda_n)$. Combined with \Cref{cor:hd-tauG-concentration}, this yields $\|\sqrt n\{\hat\tau_*(\hat B_*^{\mathrm{lasso}})-\hat\tau_*(B_*)\}\|_\infty=O_{\P}(\sqrt n\,s_n\lambda_n\rho_{G,n})$. Thus, if $\sqrt n\,s_n\lambda_n\rho_{G,n}\to0$, the Lasso-adjusted estimator is asymptotically equivalent to its oracle counterpart. The rate $\rho_{V,n}$ is also used below for plug-in covariance estimation and inference.

\begin{theorem}
\label{thm:hd-lasso-clt}
Suppose Assumptions~\ref{asu1}--\ref{asu3}, \ref{ass:hd-joint-ani}--\ref{ass:hd-penalty}, and \ref{ass:hd-oracle-clt} hold. Then, for $*\in\{\HT,\Haj\}$, $\{V_*^{\mathrm{adj}}(B_*)\}^{-1/2}\sqrt n(\hat\tau_*^{\mathrm{lasso}}-\tau)\rightsquigarrow\mathcal N(0,I_F).$
\end{theorem}

\Cref{thm:hd-lasso-clt} shows that Lasso adjustment preserves standard multivariate Gaussian inference despite the diverging dimension of the auxiliary variables. The essential requirement is not a high-dimensional Gaussian approximation for $\hat\tau_{*,G}$, but sufficiently accurate estimation of the sparse projection together with a fixed-dimensional CLT for the oracle-adjusted error. The key remainder condition can be written as $\sqrt n\,s_n(\kappa_n\rho_{V,n}+\eta_n)\rho_{G,n}\to0$. If $\kappa_n=O(1)$, $\eta_n\lesssim\rho_{V,n}$, and both $\rho_{G,n}$ and $\rho_{V,n}$ achieve the standard rate under independence, $\sqrt{\log L_G/n}$, this reduces to $s_n\log L_G/\sqrt n\to0$. Under network dependence, the two rates additionally reflect neighborhood growth, ANI decay, local-pair dependence, and the HAC bandwidth.

To conduct inference, we estimate the covariance of the Lasso-adjusted estimator using the same projection formula as in Section~\ref{sec:regression-adjustment}, $\hat V_*^{\mathrm{lasso}}=(I_F,-\hat B_*^{\mathrm{lasso}})\hat V_*(I_F,-\hat B_*^{\mathrm{lasso}})^\top.$
As in the fixed-dimensional case, the auxiliary-related covariance blocks are consistently estimable, whereas the outcome--outcome block can contain a finite-population centering bias. Let $\mathcal B_{*,\tau\tau}(b_n)=Q\mathcal B_*(b_n)Q^\top$ denote its contrast-level version. The following result shows that estimating the high-dimensional Lasso adjustment coefficients does not introduce an additional  covariance error.

\begin{theorem}
\label{thm:hd-lasso-variance}
Under the conditions of \Cref{thm:hd-lasso-clt} and  Assumptions~\ref{asu6} and ~\ref{ass:hd-variance}, $\|\hat V_*^{\mathrm{lasso}}-V_*^{\mathrm{adj}}(B_*)-\mathcal B_{*,\tau\tau}(b_n)\|_{\max}=o_{\P}(1)$.
\end{theorem}

\Cref{thm:hd-lasso-variance} separates the two sources of uncertainty in inference. The sampling covariance of the oracle-adjusted estimator is $V_*^{\mathrm{adj}}(B_*)$, while the additional deterministic term $\mathcal B_{*,\tau\tau}(b_n)$ arises from finite-population centering in the outcome covariance estimator. Thus, if $\|\mathcal B_{*,\tau\tau}(b_n)\|_{\max}\to0$, $\hat V_*^{\mathrm{lasso}}$ is consistent for $V_*^{\mathrm{adj}}(B_*)$. If instead $\mathcal B_{*,\tau\tau}(b_n)\to \mathcal B_{*,\tau\tau}^\infty\succeq0$ , the resulting inference is asymptotically conservative. Specifically, suppose $V_*^{\mathrm{adj}}(B_*)\to V^\infty\succ0$ and $\mathcal B_{*,\tau\tau}(b_n)\to \mathcal B_{*,\tau\tau}^\infty\succeq0$. An asymptotically conservative $(1-\alpha)$ confidence interval for $\tau_f$ is $\hat\tau_{*,f}^{\mathrm{lasso}}\pm z_{1-\alpha/2}\{(\hat V_*^{\mathrm{lasso}})_{ff}/n\}^{1/2}$, and an asymptotically conservative joint confidence region for $\tau$ is
\begin{equation}\label{eq:hd-lasso-confidence-region}
\left\{\theta:n(\hat\tau_*^{\mathrm{lasso}}-\theta)^\top(\hat V_*^{\mathrm{lasso}})^{-1}(\hat\tau_*^{\mathrm{lasso}}-\theta)\le\chi^2_{F,1-\alpha}\right\}. \nonumber
\end{equation}
If the centering bias is not positive semidefinite, we use the eigenvalue-adjusted covariance estimator of \citet{gao2025causal} to obtain asymptotically valid inference.

\subsection{Rerandomization and Lasso}

The preceding results consider Lasso adjustment under the original assignment mechanism. In practice, researchers may  wish to enforce design-stage balance on a small subset of  important auxiliary variables \citep{junior2025does}. Beyond precision gains, rerandomization provides an objective design-based guarantee. It mitigates concerns over p-hacking and selective analysis choices in high-dimensional settings \citep{lu2025rerandomization}. This motivates us to study rerandomization based on a fixed-dimensional subset of the auxiliary variables, followed by Lasso adjustment using the full set of auxiliary variables.

Let $q$ be fixed and let $S_n\in\mathbb R^{q\times p_n}$ be a coordinate-selection matrix whose rows are distinct standard basis vectors of $\mathbb R^{p_n}$, selecting $q$ important auxiliary contrasts to be balanced. Define $\tau_H=S_n\tau_G$ and $\hat\tau_{*,H}-\tau_H=S_n(\hat\tau_{*,G}-\tau_G)$. The corresponding covariance blocks are $V_{*,\tau H}=V_{*,\tau G}S_n^\top$, $V_{*,GH}=V_{*,GG}S_n^\top$, and $V_{*,HH}=S_nV_{*,GG}S_n^\top$. Rerandomization accepts assignments satisfying $M_{*,H}=n(\hat\tau_{*,H}-\tau_H)^\top
V_{*,HH}^{-1}
(\hat\tau_{*,H}-\tau_H)
\le a,$
where $a>0$ is fixed. As in Section~\ref{sec:rerandomization}, the balance target $\tau_H$ and the covariance entering this criterion are design-known and can be evaluated analytically or by Monte Carlo.

We define the oracle-adjusted error as the target-effect estimation error minus its population linear projection onto the auxiliary-effect estimation errors using the oracle coefficient matrix $B_*$. Under the sparsity conditions, the covariance between this error and the selected balance vector, given by $V_{*,\tau H}-B_*V_{*,GH}$, is asymptotically negligible. The oracle-adjusted error is asymptotically orthogonal to the fixed-dimensional balance variables. Thus, conditioning on the acceptance event does not alter the asymptotic distribution. These results are summarized in \Cref{thm:hd-rem-lasso} below. Let $\hat V_{*,\mathrm{ReM+Lasso}}=\hat V_*^{\mathrm{lasso}}$.

\begin{theorem}
\label{thm:hd-rem-lasso}
Suppose that the conditions of \Cref{thm:hd-lasso-clt,thm:hd-lasso-variance} hold. Suppose further that Assumption~\ref{ass:hd-oracle-clt} holds in its fixed-$q>0$ specialization, $\inf_n\lambda_{\min}(V_{*,HH})>0$, and $\P(M_{*,H}\le a)\to\P(\chi_q^2\le a)>0$. Then, for $*\in\{\HT,\Haj\}$,
\begin{equation}\label{eq:hd-rem-lasso-clt}
\left\{
\{V_*^{\mathrm{adj}}(B_*)\}^{-1/2}
\sqrt n(\hat\tau_*^{\mathrm{lasso}}-\tau)
\mid M_{*,H}\le a
\right\}
\rightsquigarrow
\mathcal N(0,I_F).
\end{equation}
Moreover,
$ \| \hat V_{*,\mathrm{ReM+Lasso}} - V_*^{\mathrm{adj}}(B_*)
- \mathcal B_{*,\tau\tau}(b_n) \|_{\max} =o_{\P}(1)$
conditionally on $M_{*,H}\le a$. If $V_*^{\mathrm{adj}}(B_*)\to V^\infty\succ0$ and $\mathcal B_{*,\tau\tau}(b_n)\to \mathcal B_{*,\tau\tau}^\infty\succeq0$, the marginal confidence intervals and joint confidence region based on $\hat V_*^{\mathrm{lasso}}$ have asymptotic coverage at least $1-\alpha$. 
\end{theorem}

\Cref{thm:hd-rem-lasso} establishes the asymptotic properties of the Lasso-adjusted estimator under rerandomization. ReM+Lasso and Lasso alone have the same asymptotic distribution. Because the Lasso adjustment removes the variation spanned by the full set of auxiliary variables, rerandomization on a fixed subset provides no further asymptotic variance reduction. Moreover, standard multivariate normal inference and conservative covariance estimation remain valid under rerandomization. Nevertheless, rerandomization can enhance finite-sample performance through improved covariate balance and higher-order properties.

\section{Simulation studies}
\label{sec:Numerical Illustration}

In this section, we evaluate the finite-sample performance of the proposed methods. We consider a random geometric (RG) network with initial size $N=1000$. Nodes are drawn independently and uniformly on $[0,1]^2$, and two nodes are connected when their Euclidean distance is at most $\{1.5/(\pi N)\}^{1/2}$. After removing isolated nodes, the retained sample contains $n=753$ units. Treatment is independently assigned with probability $1/2$. We define the exposure as $T_i=(D_i,\mathbf 1\{\sum_jA_{ij}D_j>\frac{1}{2}\sum_jA_{ij}\})$ and jointly estimate the direct and spillover (indirect) effects.

Let $\mathsf W$ denote the row-normalized adjacency matrix. For each retained sample, we generate $L_x=\lfloor0.6n\rfloor=451$ baseline covariates $X_i\stackrel{\mathrm{iid}}{\sim}N(0,\Sigma_x)$, where $(\Sigma_x)_{kl}=0.3^{|k-l|}$, $k,l=1,\ldots,L_x$. Define $G_i(D)=\{(\mathsf WD)_i,X_i^\top,(\mathsf WX)_i^\top\}^\top$. The resulting auxiliary-contrast dimension is $p_n=2(1+2L_x)=1806$, which exceeds the sample size. Potential outcomes are generated as $Y_i(D)=1.5(\mathsf WD)_i+D_i+X_i^\top\beta+D_iX_i^\top\gamma+0.5\varepsilon_i$, where $\varepsilon_i\stackrel{\mathrm{iid}}{\sim}N(0,1)$, $\beta_{1:6}=(2,-2,1.5,-1.5,1.2,-1.2)^\top$, $\gamma_{1:6}=(2.5,-2.5,2,-2,1.5,-1.5)^\top$, and all remaining coordinates are zero. The network, baseline covariates, and errors $\varepsilon_i$ are generated once and held fixed. We then generate 1,000 independent treatment assignments and report the empirical bias, standard deviation (SD), root mean squared error (RMSE), average 95\% confidence interval length (Length), and empirical coverage probability (CP).

We compare the unadjusted HT and H\'ajek estimators, two joint-contrast adaptations of the regression-adjusted H\'ajek estimators of \citet{gao2025causal} (Gao--F and Gao--L), four   joint-contrast adaptations of the covariate-adjusted estimators of \citet{lu2024adjusting} (Lu ND--F, Lu ND--$\phi_0(G_1)$, Lu ND--L, and Lu ND--$\phi_0(G_2)$), and our proposed low-dimensional (LowDim) and Lasso-adjusted (Lasso) H\'ajek estimators, both with and without rerandomization (ReM). Gao--F and Gao--L use $X_{i1}$ with common and exposure-specific slopes, respectively. The Lu specifications use $X_{i1}$ and $(\mathsf WX_1)_i$, with Lu ND--L forming exposure-specific interactions after centering these variables over the full sample and the $\phi_0(G_1)$ and $\phi_0(G_2)$ specifications additionally incorporating $(\mathsf WD)_i$. LowDim uses $\{(\mathsf WD)_i,X_{i1},(\mathsf WX_1)_i\}$, whereas Lasso uses the full vector $G_i(D)$. ReM balances the six direct- and spillover-effect contrasts generated by the treated-neighbor share, the first individual covariate, and its neighbor average. The threshold $a$ is chosen to achieve an acceptance probability of $1\%$. The Lasso penalty is selected by graph-aware five-fold cross-validation. All methods use the same positive semidefinite (PSD) projected network-HAC covariance estimator \citep{gao2025causal} with bandwidth $b_n=2$. The main text focuses on the H\'ajek results, with unadjusted HT retained as the benchmark.

Table~\ref{tab:highdim-combined} reports the results. All methods exhibit small empirical bias, and the unadjusted H\'ajek estimator is modestly more precise than HT. Relative to unadjusted H\'ajek, LowDim reduces the SD by 13.0\% and 11.9\% for the direct and spillover effects, while Lasso reduces it by 51.7\% and 42.8\%, respectively. LowDim performs comparably to the other regression-adjusted methods. No method uniformly dominates the others. ReM+Lasso attains the smallest RMSE and shortest intervals for both effects. Because the rerandomization variables are included in the high-dimensional auxiliary set, the theory predicts the same asymptotic covariance for Lasso and ReM+Lasso. Their finite-sample differences reflect higher-order effects. Finally, coverage is generally close to the nominal level. Overall, the simulation supports the proposed high-dimensional adjustment when the auxiliary dimension exceeds the sample size and shows that it can be combined with rerandomization to improve estimation efficiency without compromising inferential performance.

\begin{table}[!htb]
\centering
\caption{Simulation results under Bernoulli randomization and rerandomization ($n=753$)}
\label{tab:highdim-combined}
\begin{minipage}{\textwidth}
\setlength{\tabcolsep}{3.5pt}
\resizebox{\textwidth}{!}{%
\begin{tabular}{llrrrrrrrrrr}
\toprule
& & \multicolumn{5}{c}{Direct effect} & \multicolumn{5}{c}{Spillover effect} \\
\cmidrule(lr){3-7} \cmidrule(lr){8-12}
Design & Method & Bias & SD & RMSE & Length & CP & Bias & SD & RMSE & Length & CP \\
\midrule
Bernoulli & HT & 0.0084 & 0.4067 & 0.4066 & 1.7874 & 0.971 & -0.0089 & 0.4532 & 0.4530 & 1.8006 & 0.943 \\
Bernoulli & H\'ajek & 0.0115 & 0.3859 & 0.3859 & 1.6911 & 0.971 & -0.0068 & 0.4336 & 0.4334 & 1.6930 & 0.949 \\
Bernoulli & Gao--F & 0.0076 & 0.3338 & 0.3337 & 1.4950 & 0.971 & -0.0087 & 0.3779 & 0.3778 & 1.4989 & 0.961 \\
Bernoulli & Gao--L & 0.0069 & 0.3332 & 0.3331 & 1.4665 & 0.973 & -0.0088 & 0.3728 & 0.3727 & 1.4654 & 0.951 \\
Bernoulli & Lu ND--F & 0.0061 & 0.3349 & 0.3348 & 1.4814 & 0.969 & -0.0130 & 0.3793 & 0.3794 & 1.4856 & 0.957 \\
Bernoulli & Lu ND--$\phi_0(G_1)$ & 0.0079 & 0.3348 & 0.3348 & 1.4748 & 0.966 & -0.0123 & 0.3794 & 0.3794 & 1.4819 & 0.955 \\
Bernoulli & Lu ND--L & 0.0042 & 0.3350 & 0.3349 & 1.4338 & 0.968 & -0.0103 & 0.3770 & 0.3769 & 1.4332 & 0.942 \\
Bernoulli & Lu ND--$\phi_0(G_2)$ & 0.0131 & 0.3367 & 0.3368 & 1.4339 & 0.969 & -0.0042 & 0.3782 & 0.3780 & 1.4111 & 0.941 \\
Bernoulli & LowDim & 0.0073 & 0.3356 & 0.3355 & 1.4652 & 0.965 & -0.0062 & 0.3818 & 0.3816 & 1.4703 & 0.953 \\
Bernoulli & Lasso & 0.0011 & 0.1865 & 0.1864 & 0.8408 & 0.976 & -0.0032 & 0.2481 & 0.2480 & 0.8489 & 0.924 \\
\midrule
ReM & H\'ajek & -0.0001 & 0.3430 & 0.3428 & 1.4915 & 0.970 & -0.0123 & 0.3803 & 0.3803 & 1.4951 & 0.952 \\
ReM & LowDim & 0.0018 & 0.3357 & 0.3355 & 1.4657 & 0.969 & -0.0109 & 0.3762 & 0.3762 & 1.4703 & 0.944 \\
ReM & Lasso & -0.0053 & 0.1725 & 0.1725 & 0.8365 & 0.981 & -0.0095 & 0.2316 & 0.2317 & 0.8440 & 0.940 \\
\bottomrule
\end{tabular}%
}
\par\smallskip
\footnotesize\textit{Notes:} Bias is the empirical bias, SD is the Monte Carlo standard deviation, Length is the average 95\% confidence-interval length, and CP is the empirical coverage probability. HT is the unadjusted HT estimator, and H\'ajek is the unadjusted H\'ajek estimator.\par
\end{minipage}
\end{table}

\section{Application}
\label{sec:application}

We apply the proposed methods to the anti-conflict intervention experiment of \citet{paluck2016changing}. The study surveyed students in 56 New Jersey public middle schools and measured within-school social networks before randomization by asking students to nominate up to ten peers with whom they had recently spent time. Randomization proceeded at both the school and student levels. First, half of the schools were randomly assigned to receive the intervention. In each intervention school, researchers identified a group of seed-eligible students. This group represented approximately 15\% of the school population and was stratified by gender and grade. Half of these students were randomly chosen to serve as seeds. During the school year, seed students met regularly with trained facilitators to identify common campus conflicts. Facilitators encouraged seed students to take a visible public stance against these conflicts. Specifically, seed students created anti-conflict slogans and posters, and rewarded friendly or conflict-mitigating peers with orange program wristbands. The intervention was designed to spread anti-conflict norms and behavior through peer influence in the school network. We analyze the endline indicator \texttt{WRISTOW2}, a post-intervention outcome recording whether a student reported wearing the program wristband during the week preceding the endline survey.

Following \citet{leung2022causal,gao2025causal}, we focus on the five intervention schools. Among the 320 students eligible for individual randomization in these schools, we retain those who are nonisolated in the undirected network induced by the eligible students. This pretreatment restriction yields an analysis population of 196 students, each with at least one eligible network neighbor, so that both direct and spillover exposures can be studied on the same population. Let $S_i=\mathbf 1\{\sum_jA_{ij}D_j>0\}$ indicate whether student $i$ has at least one treated network neighbor. Our analysis retains the experiment's original stratified randomization. The 320 eligible students form 20 strata of 16 students, with exactly eight treated in each stratum. Although the joint analysis is restricted to 196 nonisolated students, generalized propensity scores and design expectations of assignment-dependent auxiliary variables are computed under the original 320-student randomization. Thus, the excluded eligible students are integrated out rather than removed from the assignment mechanism. The resulting exposure probabilities are obtained from the corresponding stratum-specific hypergeometric distributions and are positive for all four exposure conditions.
We jointly estimate the equal-weight direct contrast between $D=1$ and $D=0$ and the spillover contrast between $S=1$ and $S=0$, retaining their cross-covariance. The design-known auxiliary-effect targets for assignment-invariant pretreatment variables are zero, while the conditional expected treated-neighbor shares are evaluated under the original stratified design. We generate 20,000 legal assignments under this design to compute the additional normalization quantities required by methods of \citet{lu2024adjusting}.

We consider the same ten estimators as in the simulation study, excluding the rerandomization-based methods. The Gao adjustments use gender and grade. The Lu specifications use the same gender and grade baselines, and the relevant network-dependent specifications further include the assignment-dependent treated-neighbor share. LowDim uses gender, grade, and this share. For Lasso, we use 333 auxiliary variables comprising pretreatment individual- and network-level covariates, randomization-stratum indicators, and the treated-neighbor share. This yields $p_n=666$ auxiliary contrasts for the two effects. All methods use the same PSD-projected network-HAC covariance estimator with bandwidth $b_n=3$. The Lasso penalty is selected by graph-aware five-fold cross-validation.

\begin{figure}[!htb]
\centering
\begin{subfigure}[t]{\textwidth}
\centering
 {{\setlength{\fboxrule}{0.6pt}{\includegraphics[width=\dimexpr.9\textwidth-2\fboxsep-2\fboxrule\relax]{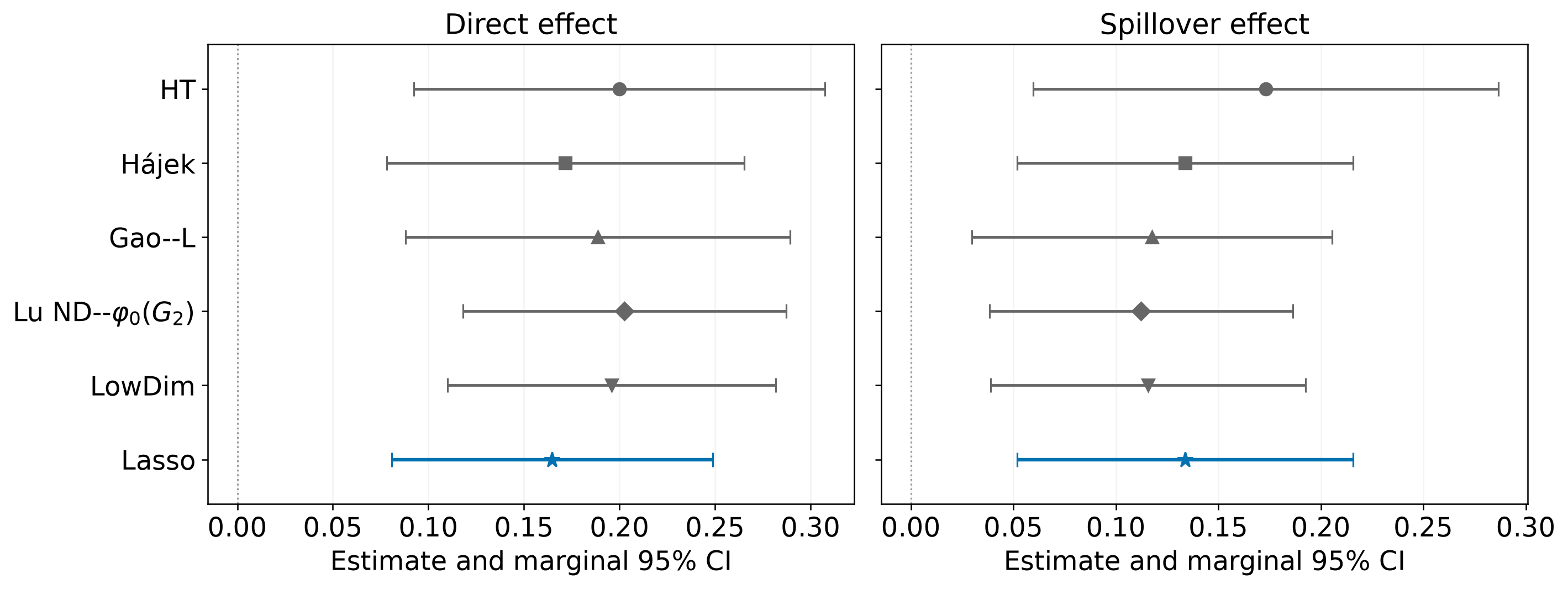}}}}
\caption{Marginal estimates and 95\% confidence intervals.}
\end{subfigure}
\vspace{0.5em}
\begin{subfigure}[t]{0.7\textwidth}
\centering
 {{\setlength{\fboxrule}{0.6pt}{\includegraphics[width=\dimexpr.68\textwidth-2\fboxsep-2\fboxrule\relax]{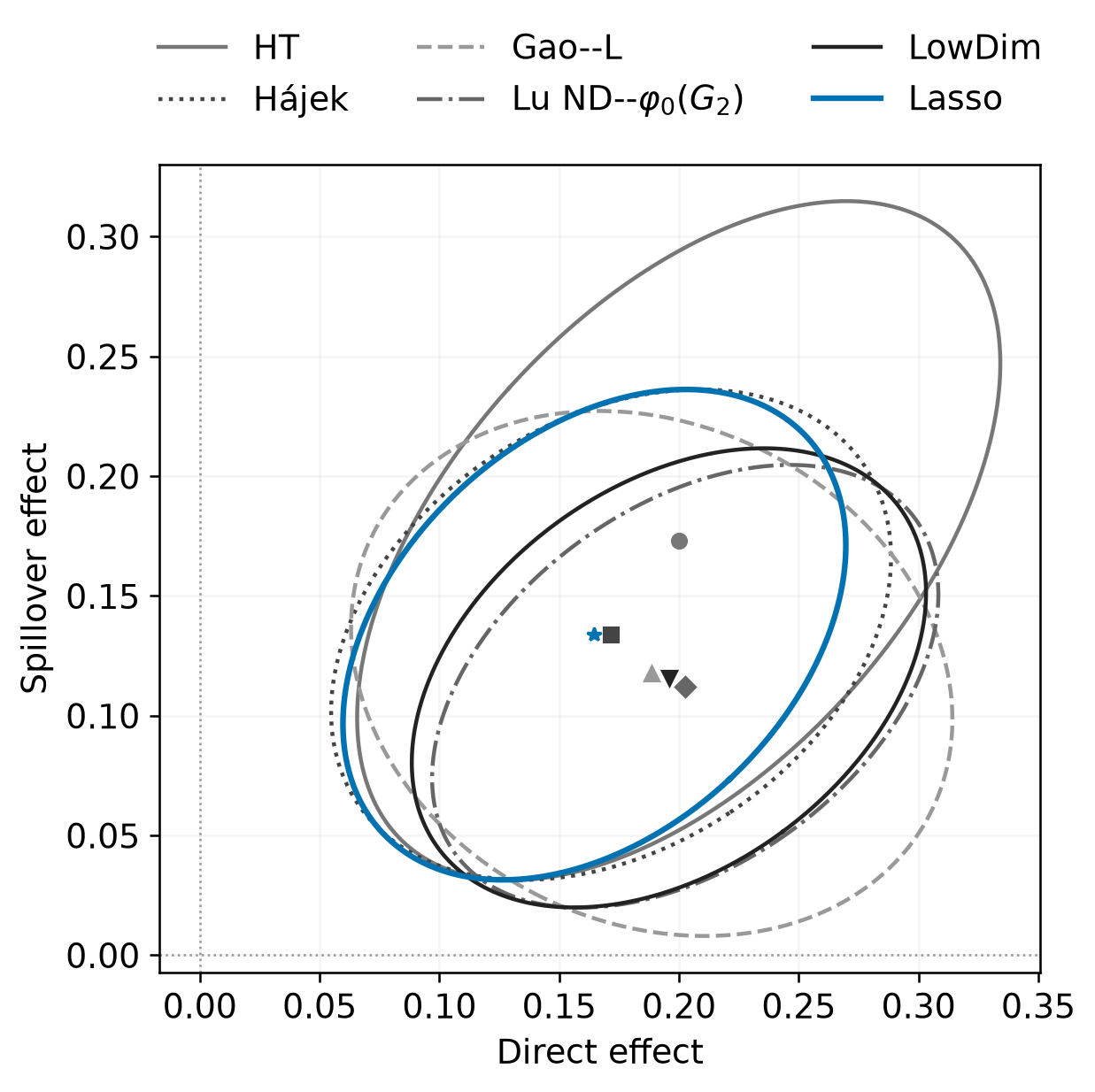}}}}
\caption{Joint 95\% confidence ellipses.}
\end{subfigure}
\caption{Joint analysis of the Paluck anti-conflict intervention on the common 196-student population. Both panels show the same six representative methods.}
\label{fig:paluck-realdata-analysis}
\end{figure}

Figure~\ref{fig:paluck-realdata-analysis} and Table~\ref{tab:paluck-realdata-analysis} show the results. Across methods, the estimated direct and spillover effects are positive and statistically significant, and the joint null of no direct or spillover effect is strongly rejected. Relative to the unadjusted H\'ajek estimator, LowDim and Lasso reduce the standard error for the direct effect by 8.2\% and  {10.2\%,} respectively. For the spillover effect, LowDim reduces the standard error by  {6.4\%,} whereas Lasso yields no reduction. Jointly, LowDim and Lasso reduce the area of the 95\% confidence ellipse by  {15.8\%} and  {11.9\%,} respectively. These results illustrate the potential precision gains from the proposed adjustment framework using both low- and high-dimensional auxiliary information.

\begin{table}[!htb]
\centering
\caption{Joint analysis of the Paluck social-network experiment}
\label{tab:paluck-realdata-analysis}
\small
\setlength{\tabcolsep}{8pt}
\begin{threeparttable}
\begin{tabular}{lcccc}
\toprule
Method & Direct effect & Spillover effect & Ellipse area & Joint $p$-value \\
\midrule
HT & 0.1999 (0.0548) & 0.1731 (0.0579) & 0.0510 & $5.78\times 10^{-4}$ \\
H\'ajek & {0.1717 (0.0477)} & 0.1338 (0.0418) & 0.0357 & $1.46\times 10^{-4}$ \\
Gao--F & 0.1782 (0.0458) & 0.1286 (0.0413) & 0.0356 & $2.24\times 10^{-6}$ \\
Gao--L & 0.1886 (0.0513) & 0.1176 (0.0448) & 0.0425 & $4.60\times 10^{-6}$ \\
Lu ND--F &  {0.1828 (0.0448)} &  {0.1314 (0.0414)} &  {0.0325} &  {$4.66\times 10^{-5}$} \\
Lu ND--$\phi_0(G_1)$ &  {0.1855 (0.0448)} &  {0.1298 (0.0407)} &  {0.0322} &  {$3.36\times 10^{-5}$} \\
Lu ND--L &  {0.1990 (0.0428)} &  {0.1185 (0.0393)} &  {0.0289} &  {$9.34\times 10^{-6}$} \\
Lu ND--$\phi_0(G_2)$ &  {0.2026 (0.0432)} &  {0.1123 (0.0378)} &  {0.0279} &  {$8.68\times 10^{-6}$} \\
LowDim &  {0.1958 (0.0438)} &  {0.1158 (0.0392)} &  {0.0300} &  {$1.73\times 10^{-5}$} \\
{Lasso} &  {0.1647 (0.0429)} & {0.1338 (0.0418)} &  {0.0314} &  {$9.63\times 10^{-5}$} \\
\bottomrule
\end{tabular}
\begin{tablenotes}
\small
\item \textit{Notes:} Effect entries are estimate (network-HAC standard error). Ellipse area is the area of the joint 95\% confidence ellipse. The last column reports the two-degree-of-freedom Wald test of a zero direct and spillover effect. 
\end{tablenotes}
\end{threeparttable}
\end{table}

The above analysis illustrates the proposed adjustments in the observed experiment, but the true effects are unknown and therefore do not permit direct evaluation of repeated-sampling accuracy or coverage. To complement this analysis, we construct a semi-synthetic experiment using the same 196-student population, network, and pretreatment auxiliary information. We specify fixed binary potential outcomes with known direct and spillover effects under the original stratified assignment mechanism. We also include predictive auxiliary variables outside the low-dimensional adjustment set to assess whether the high-dimensional Lasso can exploit additional information. We compare all ten estimators under the original-design, rerandomization (ReM), ReM+LowDim, and ReM+Lasso over 1,000 replications.

\begin{figure}[htbp]
\centering
 {{\setlength{\fboxrule}{0.6pt}{\includegraphics[width=0.9\textwidth]{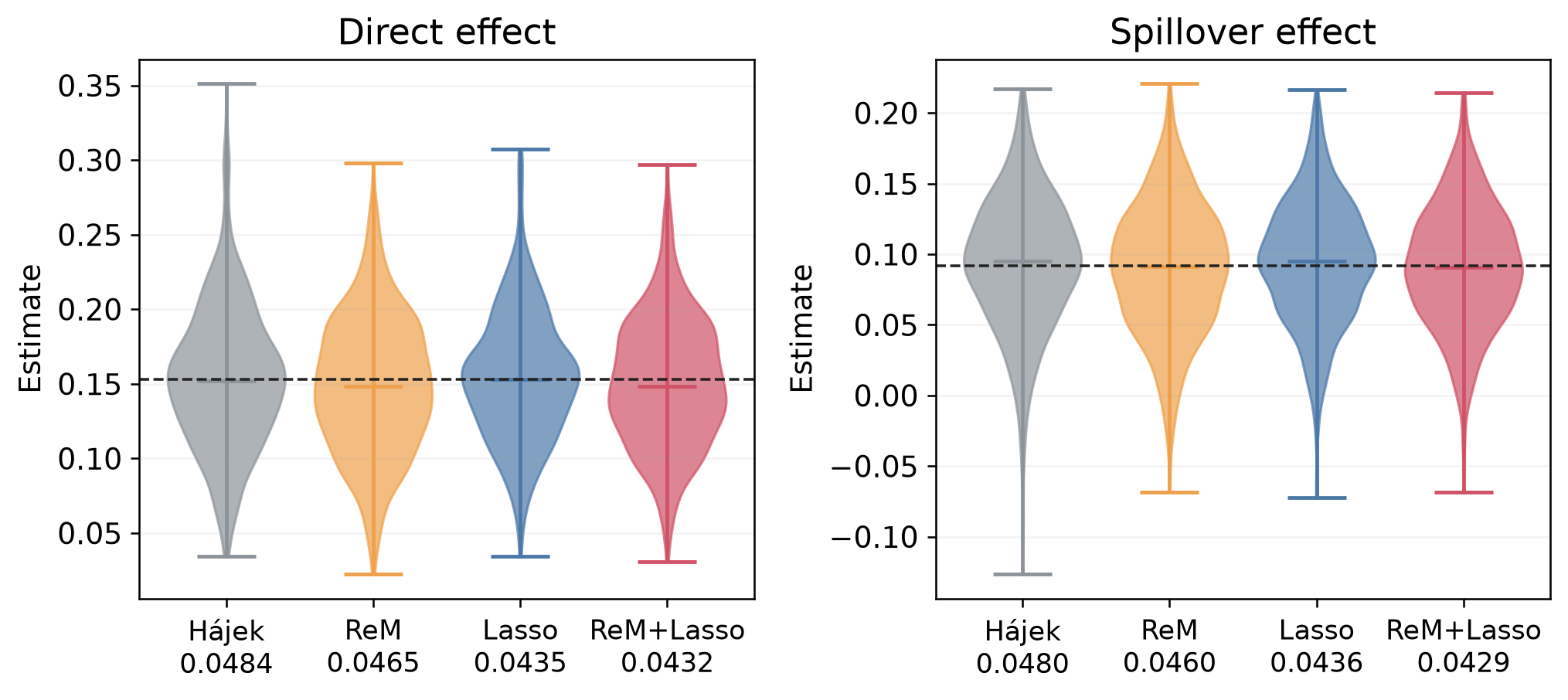}}}}
\caption{Sampling distributions in the real-data-based simulation for four representative methods. Dashed lines indicate the exact direct and spillover effects. RMSE values are reported below the method labels.}
\label{fig:paluck-realdata-simulation}
\end{figure}

Table~\ref{tab:paluck-realdata-simulation} and Figure~\ref{fig:paluck-realdata-simulation} summarize the results. All methods have small bias, while coverage ranges from 0.922 to 0.974. Relative to H\'ajek, Lasso reduces the SD by 10.3\% for the direct effect and  9.3\% for the spillover effect. ReM+Lasso yields the largest gains, reducing the SD by 11.0\% and 10.7\%, respectively, and achieves the smallest RMSE for both effects. Figure~\ref{fig:paluck-realdata-simulation} further illustrates the tighter sampling distributions of Lasso- and ReM+Lasso-adjusted estimators for both effects. These results illustrate precision gains from high-dimensional auxiliary information and its combination with rerandomization.

\begin{table}[!htb]
\centering
\captionsetup{labelfont={color=black},textfont={color=black}}
\caption{ {Semi-synthetic simulation results ($n=196$)}}
\label{tab:paluck-realdata-simulation}
 {\begin{minipage}{\textwidth}
\setlength{\tabcolsep}{3.5pt}
\resizebox{\textwidth}{!}{%
\begin{tabular}{llrrrrrrrrrr}
\toprule
& & \multicolumn{5}{c}{Direct effect} & \multicolumn{5}{c}{Spillover effect} \\
\cmidrule(lr){3-7} \cmidrule(lr){8-12}
Design & Method & Bias & SD & RMSE & Length & CP & Bias & SD & RMSE & Length & CP \\
\midrule
Stratified & HT & 0.0014 & 0.0538 & 0.0538 & 0.2321 & 0.974 & 0.0005 & 0.0573 & 0.0573 & 0.2443 & 0.974 \\
Stratified & H\'ajek & 0.0007 & 0.0485 & 0.0484 & 0.2081 & 0.969 & 0.0014 & 0.0480 & 0.0480 & 0.2069 & 0.969 \\
Stratified & Gao--F & -0.0001 & 0.0481 & 0.0480 & 0.2030 & 0.968 & 0.0008 & 0.0472 & 0.0471 & 0.2002 & 0.973 \\
Stratified & Gao--L & 0.0005 & 0.0480 & 0.0480 & 0.1926 & 0.957 & 0.0017 & 0.0466 & 0.0466 & 0.1905 & 0.951 \\
Stratified & Lu ND--F & -0.0016 & 0.0489 & 0.0489 & 0.1990 & 0.958 & 0.0000 & 0.0470 & 0.0469 & 0.1953 & 0.960 \\
Stratified & Lu ND--$\phi_0(G_1)$ & -0.0015 & 0.0495 & 0.0495 & 0.1966 & 0.952 & -0.0007 & 0.0478 & 0.0478 & 0.1915 & 0.957 \\
Stratified & Lu ND--L & 0.0007 & 0.0485 & 0.0485 & 0.1798 & 0.930 & 0.0005 & 0.0471 & 0.0471 & 0.1752 & 0.936 \\
Stratified & Lu ND--$\phi_0(G_2)$ & -0.0065 & 0.0497 & 0.0501 & 0.1743 & 0.925 & -0.0016 & 0.0492 & 0.0492 & 0.1703 & 0.922 \\
Stratified & LowDim & -0.0012 & 0.0469 & 0.0469 & 0.1864 & 0.955 & -0.0028 & 0.0460 & 0.0461 & 0.1801 & 0.943 \\
Stratified & Lasso & 0.0013 & 0.0435 & 0.0435 & 0.1787 & 0.954 & 0.0028 & 0.0436 & 0.0436 & 0.1770 & 0.945 \\
\midrule
ReM & H\'ajek & -0.0039 & 0.0464 & 0.0465 & 0.1870 & 0.949 & -0.0019 & 0.0459 & 0.0460 & 0.1812 & 0.945 \\
ReM & LowDim & -0.0039 & 0.0464 & 0.0465 & 0.1848 & 0.945 & -0.0029 & 0.0461 & 0.0461 & 0.1790 & 0.944 \\
ReM & Lasso & -0.0025 & 0.0431 & 0.0432 & 0.1749 & 0.946 & -0.0003 & 0.0429 & 0.0429 & 0.1740 & 0.944 \\
\bottomrule
\end{tabular}%
}
\endgraf\smallskip
\footnotesize\textit{Notes:} Bias, SD, RMSE, Length, and CP use the definitions in the simulation study. \endgraf
\end{minipage}}
\end{table}

\section{Discussion} \label{sec:conclusions}

In this paper, we develop rerandomization and covariate adjustment for joint exposure-effect estimation and inference under network interference. Our approach uses auxiliary-effect estimation errors, yielding an adjustment asymptotically equivalent to rerandomization under a vanishing acceptance threshold. The methodology extends to high-dimensional settings via Lasso regularization. We establish concentration bounds and asymptotic distributions for Lasso-adjusted estimators, and we provide valid inference procedures. The empirical application and semi-synthetic experiment illustrate precision gains from low- and high-dimensional adjustments, with high-dimensional Lasso exploiting richer auxiliary information.

Several directions merit further study. First, exposure-based inverse-probability weighting can be inefficient for global treatment effects under rare exposure conditions. Extending our balancing and projection methods to estimators designed for global effects \citep{eckles2017design,yu2022estimating} would be useful. Second, debiasing the high-dimensional covariate adjustment would be valuable. In practice, many auxiliary variables have small but non-negligible predictive effects. This violates strict sparsity assumptions. Standard Lasso adjustments then introduce penalization bias and invalidate inference. Constructing a debiased adjustment may eliminate this bias, weakens sparsity requirements, and restores standard asymptotic normality and valid inference \citep{lei2021regression,lu2025debiased}. Third, extending the theory beyond independent Bernoulli assignment to stratified, clustered, and covariate-adaptive randomization designs would broaden applicability to complex network experiments.

\bibliography{bibliography.bib}
\end{document}